\documentclass[aps,pre,preprint,superscriptaddress,floatfix]{revtex4-1}
\usepackage{bm}
\usepackage{graphicx,color}
\usepackage{dcolumn}
\usepackage{times}

\usepackage{amsmath,amssymb}
\usepackage{changepage}

\usepackage{textcomp,marvosym}

\usepackage{nameref,hyperref}

\usepackage[right]{lineno}

\usepackage[nopatch=eqnum]{microtype}
\DisableLigatures[f]{encoding = *, family = * }

\usepackage[table]{xcolor}

\usepackage{array}

\newcolumntype{+}{!{\vrule width 2pt}}

\newlength\savedwidth

\newcommand\thickhline{\noalign{\global\savedwidth\arrayrulewidth\global\arrayrulewidth 2pt}%
\hline
\noalign{\global\arrayrulewidth\savedwidth}}

\usepackage[aboveskip=1pt,labelfont=bf,labelsep=period,justification=raggedright,singlelinecheck=off]{caption}

\makeatletter
\renewcommand{\@biblabel}[1]{\quad#1.}
\makeatother

\usepackage{lastpage,fancyhdr,graphicx}
\usepackage{epstopdf}
\fancyheadoffset[L]{2.25in}
\fancyfootoffset[L]{2.25in}
\usepackage{tikz}
\usetikzlibrary{shapes.geometric, arrows}

\tikzstyle{startstop} = [rectangle, rounded corners, minimum width=3cm, minimum height=1cm,text centered, draw=black, fill=red!30]
\tikzstyle{process} = [rectangle, minimum width=3cm, minimum height=1cm, text centered, draw=black, fill=blue!20]
\tikzstyle{processFEM} = [rectangle, minimum width=3cm, minimum height=1cm, text centered, draw=black, fill=orange!30]
\tikzstyle{decision} = [diamond, minimum width=3cm, minimum height=1cm, text centered, draw=black, fill=green!30]
\tikzstyle{arrow} = [thick,->,>=stealth]

\begin{document}
\title{Early stages of collective cell invasion: Biomechanics}
\author{R. Gonz\'alez-Albaladejo$^{*}$}
\affiliation{Universidad Carlos III de Madrid, Gregorio Mill\'an Institute for Fluid Dynamics, Nanoscience and Industrial Mathematics, 28911 Legan\'{e}s, Spain.}
\affiliation{Universidad Carlos III de Madrid, Department of Mathematics, 28911 Legan\'{e}s, Spain.}
\affiliation{$^{*}$ORCID: 0000-0001-9560-5720}
\author{M. Carretero$^{**}$}
\affiliation{Universidad Carlos III de Madrid, Gregorio Mill\'an Institute for Fluid Dynamics, Nanoscience and Industrial Mathematics, 28911 Legan\'{e}s, Spain.}
\affiliation{Universidad Carlos III de Madrid, Department of Mathematics, 28911 Legan\'{e}s, Spain.}
\affiliation{$^{**}$ORCID: 0000-0002-3517-4241}
\author{L. L. Bonilla$^{***}$}
\affiliation{Universidad Carlos III de Madrid, Gregorio Mill\'an Institute for Fluid Dynamics, Nanoscience and Industrial Mathematics, 28911 Legan\'{e}s, Spain.}
\affiliation{Universidad Carlos III de Madrid, Department of Mathematics, 28911 Legan\'{e}s, Spain.}
\affiliation{$^{***}$Corresponding author. E-mail: bonilla@ing.uc3m.es. ORCID: 000-0002-7687-8595}
\date{\today}
\begin{abstract}
The early stages of the collective invasion may occur by single mesenchymal cells or hybrid epithelial-mesenchymal cell groups that detach from cancerous tissue. Tumors may also emit invading protrusions of epithelial cells, which could be led (or not) by a basal cell. Here we devise a novel fractional step cellular Potts model comprising passive and active cells able to describe these different types of collective invasion before cells start proliferating. Cells moving toward stiffness gradients (durotaxis) and active forces pulling them away from the tumor have different symmetry properties under cellular extension and retraction that sometimes hamper collective invasion when put together. Thus,  these forces are included in different half steps of the fractional step method. Compared with a single step method, fractional step produces more realistic collective invasion scenarios with little extra computational effort. Biochemical mechanisms that determine how cells acquire their different phenotypes and cellular proliferation will be incorporated to the model in future publications.
\end{abstract}
\maketitle


\section{Introduction}

The invasion of a tissue by cancer cells is a crucial first step of metastasis \cite{fri12,vil21}. As cells move, they interact with the extracellular matrix (ECM), which plays an important role in cancer \cite{sle24}. Consider a tumor comprising epithelial (E) cells that adhere to each other and form the cancerous tissue. There are several types of invasion into healthy tissue after cancer cells break the basal membrane surrounding the tumor: either by single cells or by groups of cells in a collective invasion \cite{fri12,sle24}. Collectively invading cells maintain cell-cell junctions, whereas individually invading cells fully detach from the rest of the tumor. Different types of invasion are as follows.

(i) E cells undergoing the epithelial-mesenchymal transition (EMT) \cite{pas18} lose their apical-basal polarity and cell-cell adhesions and reorganize their cytoskeleton, eventually obtaining a mesenchymal (M) phenotype with front-back polarity that enables migration away from the tumor as single cells \cite{vil21}. M cells are elongated, emit spindle-like (filopodia) or sheet-like (lamellipodia) protrusions and advance through adhesion-mediated tractions and ECM degradation \cite{sha20}. The migration of M cells may pave a path through the ECM for E cells to follow \cite{gag07}. In complex, heterogeneous ECM microenvironments, single cells may adopt a much faster ameboid motion (about 10$\mu$m/min vs 0.3$\mu$m/min): they become round, emit blebs or pseudopods, have short-term and weak attachments to the ECM, and can squeeze through ECM pores \cite{pan10,tal17,sha20}. Computational models of ameboid and mesenchymal migration are discussed in Ref.~\cite{sha20}. See also \cite{wur26}.

(ii) Collective invasion may occur in different ways. Hybrid epithelial/mesen\-chymal (E/M) cells are related to different stages of the EMT \cite{pas18,bra18}, and migrate as small clusters. M and E/M cells at different stages of the EMT are found in a variety of tumors \cite{pas18}. 

(iii) Cancer tissue forms protrusions comprising a number of E cells that are alike, in an instability similar to finger formation in wound healing \cite{hak17}, or 

(iv) The protrusions are led by basal cells. Basal tumor cells have E phenotype and lead invasion without undergoing the EMT \cite{cho14}. 

Biochemical cues, EMT and cellular Notch signaling may trigger the invasion and they are intimately connected to biomechanical cellular motion and interaction \cite{boc18,vil21,vil22,muk22}. 
%
%
Bocci {\em et al} have modeled EMT, cancer stem cells and Notch signaling by sets of coupled ordinary differential equations (ODEs) \cite{boc18}. They have studied phenotypes associated to each of the three cycles and explained how treatments inhibiting some variable in the cycles perform \cite{boc18}. Distinguishing phenotypes and assigning them mechanical properties is of paramount importance when using a realistic computational tool such as the cellular Potts model (CPM) to characterize cellular motion \cite{Graner,Glazier,vanOers}. For example, in early stage angiogenesis, the phenotype tip (leader), stalk (follower) or hybrid tip/stalk (leader/follower) cell depends on biochemical factors such as the content of VEGF and Delta protein in the cells. These contents are determined by the ODEs of Notch signaling dynamics \cite{boa15,jol15,boa16}. The CPM parameters measuring the phenotype sensitivity to mechanical and chemical cues, as well as the ability to proliferate, determine the behavior of the cells \cite{veg20,veg21}. It is crucial to appropriately select their values and to check their effects \cite{veg20}. Once they are known, appropriate parameter values and qualitative properties need to be assigned to cells according to their phenotype: for angiogenesis, sensitivity to chemotaxis and proliferation of tip, stalk and hybrid tip/stalk cells  \cite{veg20}. 

Collective cell invasion depends on the ECM structure. Its role in cell migration and morphogenesis is discussed in Ref.~\cite{cro24}. The interstitial ECM fills spaces between organs and it has fibrous proteins and proteoglycans that form 3D arrangements. The basement membrane ECM forms 2D sheet-like ECM that line organ boundaries \cite{cro24}. The influence of a fibrous, cross-linked ECM (modeled by bead-springs) on cell motion modeled by a CPM is discussed in Ref.~\cite{tsi23}. The  emergence of different cell motility types from cell-matrix adhesion dynamics via a modified CPM is shown in Ref.~\cite{vst22}. For collective invasion of cancer cells, the relation between biomechanics and the phenotypes within the EMT, Notch signaling and cancer stem cell cycles \cite{boc18} has to be specified. Recently, Hirway {\em et al} have incorporated EMT to the CPM of Ref.~\cite{vanOers}, either by means of a set of ODEs for one type of cells and for the ECM \cite{hir21}, or for several cell types and for the ECM \cite{hir24}. These ODEs include source terms dependent on junctional forces that are coupled to the equations of elasticity comprising the mechanical part of the model \cite{hir21}. They determine the E, M and E/M phenotype of the cells, each having different adhesion and proliferation properties. The resulting model describes tissue formation and a wound healing (scratch) assay in which the cells of a tissue close a space that has opened in their midst \cite{hir21}. In these simulations, the cells have to proliferate to complete the tissue or to close the opening. However, experiments in wound healing assays \cite{pet10} or in antagonistic migration assays \cite{moi19} show that openings between epithelial tissues may close before cells have time to proliferate. Thus, it is important to model early stages of collective invasion before cells proliferatate.

Different experiments have shown different phases in migrating clusters of malignant cells that can be modeled using ideas from active matter \cite{cop18}. Experiments, theory and numerical simulations have shown the existence of chiral currents in confined monolayers of highly active spindle-shaped human fibrosarcoma cells with sources of active forces at the interfaces \cite{yas22}. Long range force transmission of mechanical origin fosters phenomena such as waves in antagonistic migration assays \cite{rod17}, stress fibres mechanics \cite{oak17} and collective cell durotaxis \cite{sun16}. 
A CPM with migrating cells subject to active Brownian motion describes ameboid motion in a porous material \cite{wur26}. Cellular self-propulsion and inhomogeneous polarized cell-cell adhesion affect collective  migration \cite{mat17}. Cell-cell adhesion and active forces due to a two-state signaling protein inside cells may modify the CPM energy to incorporate the solutions of reaction-diffusion equations for the concentrations of active/inactive protein states \cite{ren19}. Including pseudopods in the CPM can be achieved by defining a persistence variable related to actin and adding appropriate terms to the CPM energy \cite{bur22}.

Here we focus on the mechanical aspects of cellular motion and the early stages of invasion before cells proliferate. Our work provides a proof-of-concept biomechanics framework for invasion prior to cell proliferation, distinguishing it from biochemical EMT or signaling models. Starting from a CPM that includes durotaxis \cite{vanOers}, we add {\em novel active forces} that are felt only by mesenchymal and hybrid cells, not by epithelial cells. Each step of a novel fractional step Monte Carlo dynamics is divided into two substeps: traction cells that are the basis of durotaxis intervene in the first substep whereas active forces act on the second substep. Each cell type has a specific local stiffness threshold that characterizes its response to strain gradients. E cells are static and very adhesive, M cells are motile, less adhesive and freer to move, whereas E/M hybrid cells combine both behaviors. M and hybrid E/M cells are subject to active forces that do not affect E cells. We consider a circular aggregate comprising E, M and hybrid E/M cells and study their migration and invasion at times shorter than that of cellular division. 
Our results show that mesenchymal and hybrid cells placed near the edges of the aggregate escape toward an attraction point as single cells and small aggregates, respectively. Hybrid cells are more effective in their migration. Biochemical cues that include coupled EMT, cancer stem cell and Notch signaling dynamics will be incorporated in a subsequent paper. Our work is a proof of concept that a combination of cellular phenotypes and mechanical properties can model early individual and collective invasion of tissues with possible applications for tumors and cancer metastasis. 

The rest of the paper is organized as follows. Section \ref{sec:2} reviews the CPM with durotaxis. Section \ref{sec:3} explains how cell adhesion and a threshold of the Young modulus affect cellular dynamics and distinguish passive and motile cells. Section \ref{sec:4} introduces active migration forces for motile cells and argues that they need to be considered separately in a fractional step CPM to achieve effective migration. Section \ref{sec:5} describes the different types of collective cellular invasion and the obtained results are discussed in Section \ref{sec:6}. In Appendix \ref{ap:a}, starting from a circular aggregate of passive and motile cells, we enumerate 21 configurations and parameter values that produce seven different patterns. Appendix \ref{ap:b} contains flowcharts of single step and fractional step CPMs.

\section{Dynamics of Cells on a Substrate: Durotaxis}\label{sec:2}
Epithelial cells have a propensity to migrate up gradients of substrate rigidity, which is called durotaxis \cite{vanOers,lo00}. In this section, we review van Oers {\em et al}'s CPM with durotaxis, which is calibrated with data from experiments \cite{vanOers}. 
Thus, we will  
assume that a strained ECM is stiffer along the strain orientation than perpendicular to it (strain stiffening), that the ECM is isotropic and linearly elastic, and that stiffness is an increasing, linear function of the local strain. Even for homogeneous ECM, strains due to the traction of the cells generate stiffness gradients by modifying the elastic constants (e.g., the Young modulus). 


We solve the Navier equations for isotropic linear elasticity using the finite element method (FEM)  as follows:
\begin{eqnarray}
\mathbf{K}\mathbf{u} &=& \mathbf{f}, \quad \mathbf{f}_i = \mu\sum_j \mathbf{d}_{ij},\quad \mathbf{d}_{ij} = \mathbf{n}_j - \mathbf{n}_i. \label{eq1}
\end{eqnarray}
The FEM represents the substrate as a lattice of finite elements $e$, each corresponding to a square pixel of the CPM having four nodes. In Eq.~\eqref{eq1}, $\mathbf{K}$ is the stiffness matrix, $\mathbf{u}$ the displacement vector, and $\mathbf{f}$ the force vector. The vector $\mathbf{u}= (u_{x_1},u_{y_1},\ldots,u_{x_n},u_{y_n})^T$ contains the displacements of all nodes, which are the unknowns that the FEM calculates based on the traction forces $\mathbf{f}$ exerted by the cells onto the material. Force $\mathbf{f}_i$ at node $i$ (with position $\mathbf{n}_i$) is the tension per unit length, $\mu$, multiplied by the sum of distances $\mathbf{d}_{ij}$ between node $i$ and all other nodes nodes $j$ at $\mathbf{n}_j$ within the same cell. The force $\mathbf{f}_i$ mimics the cell-shape dependent contractile forces exerted by the cells onto the ECM, it does not derive from a potential energy, and it gives experimentally plausible predictions for fibroblasts, endothelial cells, and keratocytes \cite{lem10}.

In the present CPM, the pixels $\mathbf{x}$ of a square grid (or in a cube in 3D) have labels identifying them as part of a cell or the ECM. After a random pixel change in which the label of the source $\mathbf{x}$  is copied to the target $\mathbf{x'}$, the CPM uses the Metropolis algorithm to update an energy $H$ defined on the pixels. The energy comprises a term enforcing cell volume, a contact term modeling cell-cell and cell-substrate adhesion and a term monitoring changes in strain. The probability of accepting a change $\Delta H$ in energy is 1 if $\Delta H<0$ and $e^{-\Delta H/T}$ otherwise:
\begin{eqnarray}
P(\Delta H) &=& \begin{cases} 1 & \text{if } \Delta H < 0 \\ e^{-\Delta H/T} & \text{if } \Delta H \geq 0 \end{cases}, \label{eq2}
\end{eqnarray}
where the temperature parameter $T > 0$ controls the intrinsic cell motility. The change in energy (Hamiltonian) is 
\begin{eqnarray}
&&\Delta H = \Delta H_\text{volume}+\Delta H_\text{contact}+\Delta H_\text{duro}, \label{eq3}\\
&&H_\text{volume} = \sum_{\sigma\in cells} \lambda \left(\frac{a(\sigma)-A(\sigma)}{A(\sigma)}\right)^2, \label{eq4}\\
&&H_\text{contact}= \sum_{\mathbf{x},\mathbf{x}'} J(\sigma(\mathbf{x}),\sigma(\mathbf{x}'))\left(1-\delta(\sigma(\mathbf{x}),\sigma(\mathbf{x}'))\right). \label{eq5}
\end{eqnarray}
Here $a(\sigma)$, $A(\sigma)$, $\sigma$, $\lambda$, and $\delta(.,.)$ are the actual cell volume, target volume, an integer representing the label of each cell and substrate, the strength of the volume constraint, and the Kronecker delta, respectively. $J(\sigma(\mathbf{x}), \sigma(\mathbf{x}')) \geq 0$ is the adhesion strength between neighboring pixels $\mathbf{x}$ and $\mathbf{x}'$, which is larger for cell-cell interactions than for cell-substrate (e.g., $J_{cc} \approx 2J_{cs}=2.5$). Adding a perimeter constraint to Eq.~\eqref{eq4} scarcely influences CPM simulation results \cite{veg20} and in its effect can be often subsumed in the adhesion parameter \cite{dur19}.

The strain term in Eq.~\eqref{eq3} captures cell responses to substrate stiffness gradients (durotaxis) \cite{vanOers}:
\begin{eqnarray}
\Delta H_\text{duro} &=& -g(\mathbf{x},\mathbf{x}')\lambda_{\text{duro}}\left[h(E(\epsilon_1))(\mathbf{v}_1\cdot\mathbf{v}_m)^2+h(E(\epsilon_2))(\mathbf{v}_2\cdot\mathbf{v}_m)^2\right]\!. \label{eq6}
\end{eqnarray}
Here $g(\mathbf{x}, \mathbf{x}')$ distinguishes extension ($+1$) and retraction ($-1$), $\lambda_{\text{duro}}$ controls the strain contribution, and $\mathbf{v}_m$ is a unit vector in the copy direction $\mathbf{x}' - \mathbf{x}$. The eigenvalues $\epsilon_1$ and $\epsilon_2$ of the strain tensor are the principal strains, and the corresponding eigenvectors, $\mathbf{v}_1$ and $\mathbf{v}_2$, provide the strain orientation. Writing now the strain tensor as a column vector, we have
\begin{eqnarray}
\mathbf{\varepsilon} = \mathbf{B}\mathbf{u}_e, \label{eq7}
\end{eqnarray}
where $\mathbf{B}$ is the conventional strain-displacement matrix for a four-noded quadrilateral element, and $\mathbf{u}_e$ are the node displacements related to the local strains. 

In Eq.~\eqref{eq6}, the substrate stiffness measured by the ECM Young modulus, $E(\epsilon)$, depends linearly on the principal strains:
\begin{eqnarray}
E(\epsilon) = E_0\!\left[1+\frac{\epsilon}{\epsilon_{\text{st}}}\delta(|\epsilon|,\epsilon)\right]\!, \label{eq8}
\end{eqnarray}
where $\epsilon_{\text{st}}$ adjusts stiffening (which only occurs for extensions $\epsilon>0$, not compressions $\epsilon<0$). The sigmoidal function $h(E)$ regulates focal adhesion maturation based on the Young modulus threshold (YMT) $E_\theta$:
\begin{eqnarray}
h(E)&=&\frac{1}{1 + e^{-\beta(E-E_{\theta})}}, \label{eq9}
\end{eqnarray}
where $\beta$ is the steepness of the sigmoid. In Eq.~\eqref{eq6}, we use the strain value of the target pixel $\mathbf{x}'$ in extension, and that of the source pixel $\mathbf{x}$ in retraction. This extension-retraction symmetry keeps the cohesion of the cells. The force vector in Eq.~\eqref{eq1} changes the energy Eq.~\eqref{eq6} through the strain tensor that also modifies the substrate stiffness. The strain-driven energy change entering the Monte Carlo time step (MCTS) is the effective force moving the cells.

The two-dimensional computational domain (substrate) is defined by its size in pixels, with $NV_x$ and $NV_y$ representing the number of pixels along each side. Each pixel corresponds to a physical size of $dx=2.5\mu m$ resulting in domain dimensions $Lx = NV_x \cdot dx$ and $Ly = NV_y \cdot dx$. In this section, we set $NV_x = NV_y = 300$, which gives a domain size of $750 \mu$m $\times$ $750 \mu$m. The parameter values listed in Table \ref{t1} are compatible with those in Ref.~\cite{vanOers}, capture the qualitative behavior of the system and ensure numerical stability and consistency with experimental observations. The reference adhesion parameter $J_\text{ref}$ is representative of the adhesion moduli for the different cell phenotypes.
\begin{table}[!ht]
		\caption{
			{\bf  Parameter values.}}
		\begin{tabular}{r|c|c|c|c|c|c|c|c|c}
			\text{Parameter} & $\lambda$ &  $\lambda_\text{duro}$ & $T$ & $\epsilon_\text{st}$ & $J_\text{ref}$& $\mu$& $E_0$ &$E_\theta$& $\beta$\\ \hline
			\text{Value}	&  500    & 10  & 1 & 0.1 & 1.25 & 0.01 & 10 & 1-15& 0.5\\ \hline
			\text{Unit} & - & -& - & - & - & nN/$\mu$m & kPa & kPa & kPa$^{-1}$
		\end{tabular}
		\label{t1}
\end{table}

\section{Young modulus threshold and adhesion in durotaxis}\label{sec:3}
Here we consider the behavior of two types of cells that differ in their YMT and/or adhesion properties under the CPM Eqs.~\eqref{eq1}-\eqref{eq9}. One of the cells will turn out to be passive and remain epithelial cells sharing line boundaries whereas the others will be motile cells that may even leave the tumor. How do we proceed?

The interplay between strain and substrate stiffness creates a feedback loop that governs cell dynamics and collective behavior. When cells adhere to a deformable substrate, they exert traction forces that deform it. The deformation changes the substrate's stiffness, which in turn influences cell migration and adhesion dynamics. This feedback loop is governed by the strain-stiffening relationship in Eqs~(\ref{eq7})-(\ref{eq8}). The strain-induced changes in stiffness affect the Hamiltonian strain term through Eqs.~\eqref{eq6} and (\ref{eq9}). In extension, we use the YMT of the target pixel $\mathbf{x}'$, whereas we use the YMT of the source pixel $\mathbf{x}$ in retraction. Cells with a large YMT produce large strains in extension and move more frequently according to Eq.~\eqref{eq8}-\eqref{eq9}. Motile cells exhibit weak interactions, while passive cells with a smaller YMT cluster tightly. A more elaborated CPM  explains durotaxis from cell-ECM forces and focal adhesion dynamics \cite{ren20}.

\begin{figure}[!h]
\includegraphics[clip,width=4.3cm]{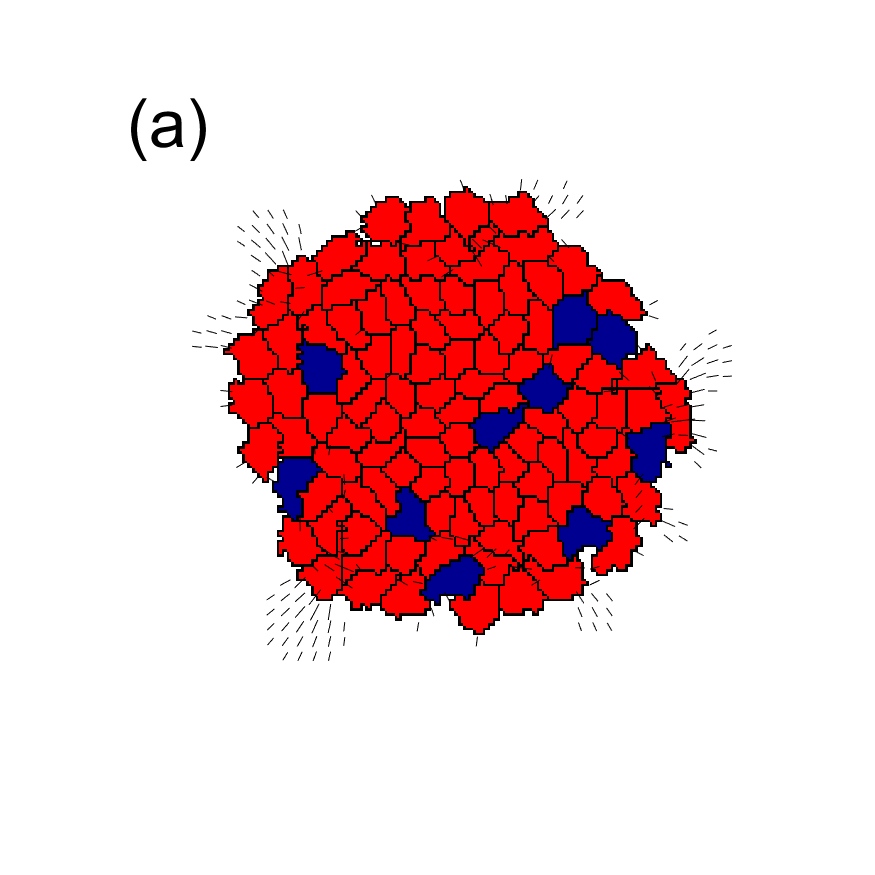}
\includegraphics[clip,width=4.3cm]{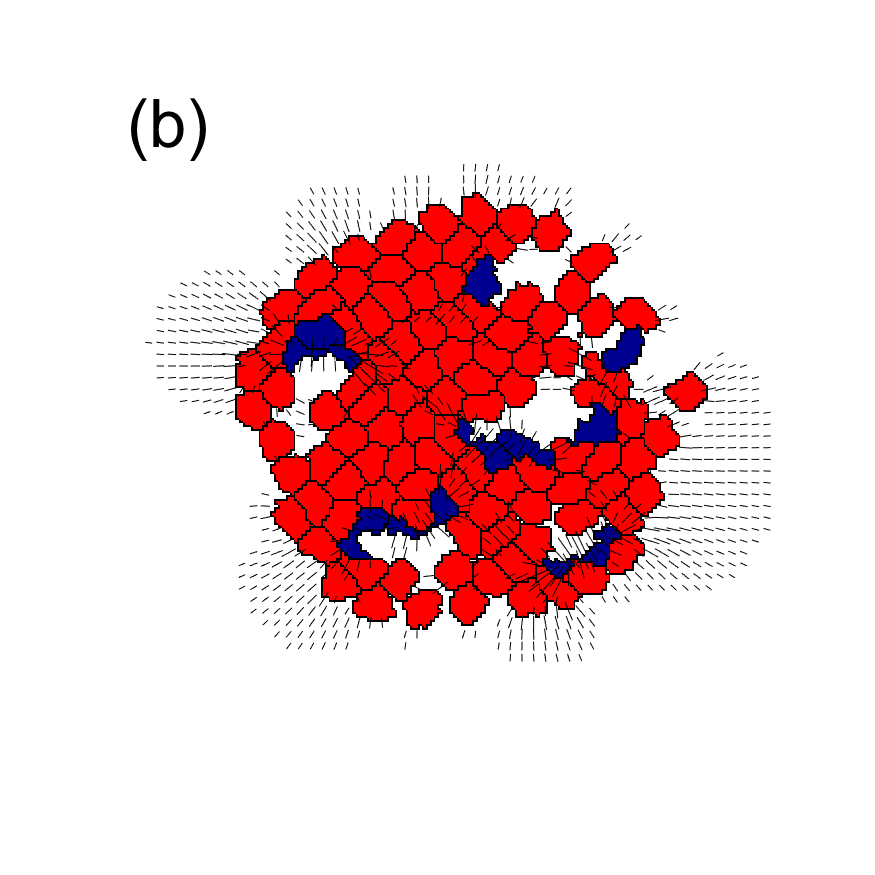}
\includegraphics[clip,width=4.3cm]{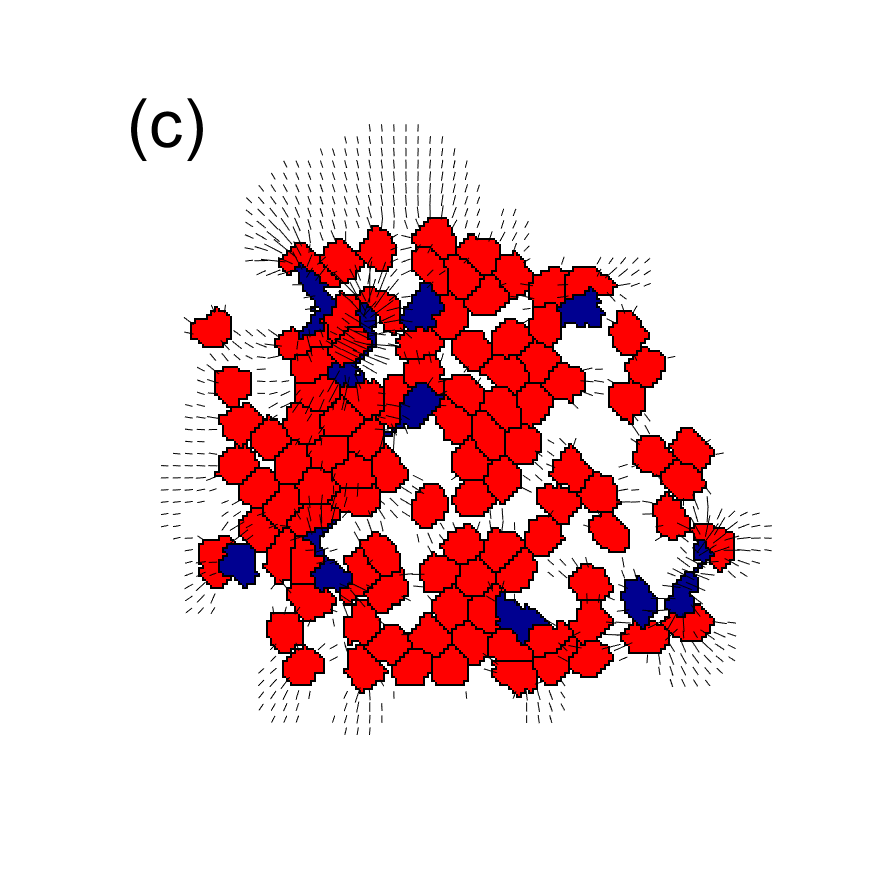}
\caption{{\bf Dynamics of motile and passive cells having different YMTs $E_\theta$.} Initially,
10 blue motile cells ($E_{\theta m}=15$ kPa) are randomly embedded into a disk of 101 passive red cells ($E_{\theta c}=1$ kPa) with relative adhesion values $j_{cc}=j_{cm}=j_{mm}=2$, $j_{cs}=j_{ms}=1$, where $J_{kl}=j_{kl}J_\text{ref}$ (see Subsection \ref{ssec:2_2}). Black segments indicate strain magnitude and orientation above a given threshold \cite{vanOers}. Monte Carlo time steps (MCTS): (a) t=30 , (b) t=250, (c) t=500. See Video S1 \cite{suppl}. } 
\label{fig1}
\end{figure}

\subsection{Effects of the Young modulus threshold} \label{ssec:2_1}
By means of Eqs~(\ref{eq1})-(\ref{eq9}), we have simulated the evolution of a circular agglomerate of cells with different YMTs: $E_{\theta c} = 1$ kPa  (red non-motile cells) and $E_{\theta m} = 15$ kPa  (blue motile cells). Non-motile cells maintain equilibrium under substrate rigidity, while motile cells respond to strain gradients generated by cells of the other type. The result is that motile cells proceed towards the borders of the agglomerate, as shown in Fig~\ref{fig1}. The aggregate breaks down in different ways depending on the initial distribution of cells. See Video S1 \cite{suppl}.

\subsection{Patterns due to adhesion and YMT}\label{ssec:2_2}
To incorporate a hybrid cell type that exhibits both motile behavior and a tendency to adhere to similar cells, we modify the cell-cell and cell-substrate adhesion parameters, $J(\sigma(\mathbf{x}), \sigma(\mathbf{x}'))$, in Eq~(\ref{eq5}). Their values influence the interactions between various cell types and the substrate, thereby allowing to identify cellular phases with distinct mechanical properties as phenotypes. The latter may result from cell biochemical processes (which we discuss in a forthcoming paper). In a similar vein, different mechanical properties are assigned to leader and follower (tip and stalk) cells resulting from Notch signaling dynamics in angiogenesis \cite{veg20}.

\begin{figure}[!h]
\includegraphics[clip,width=4.3cm]{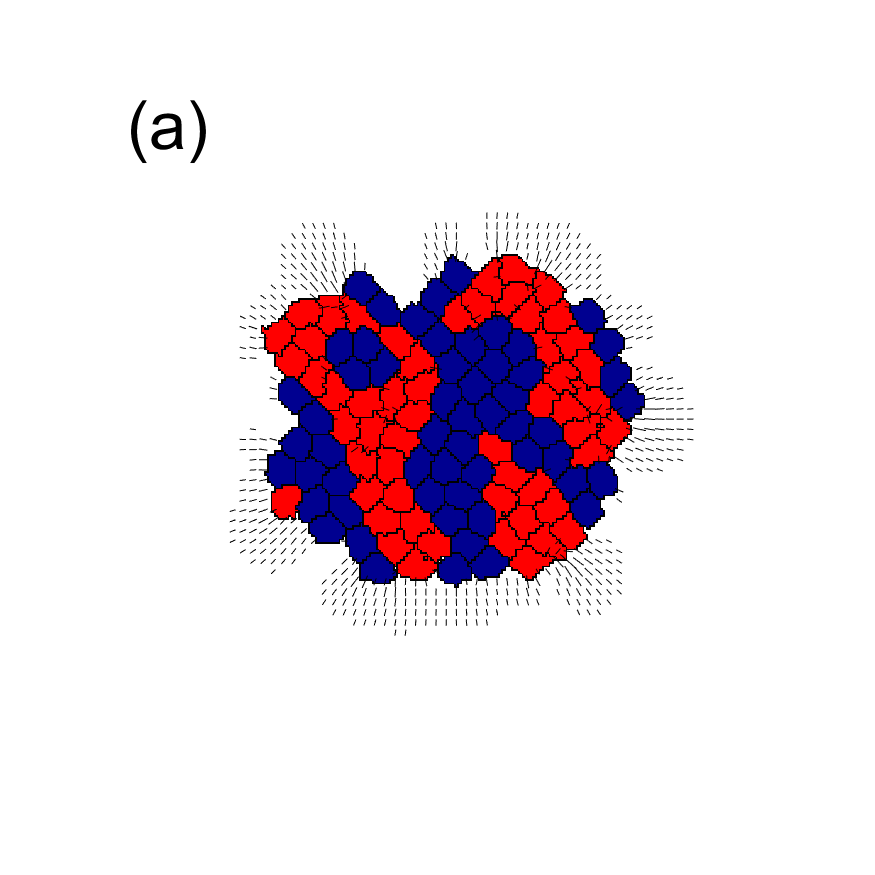}
\includegraphics[clip,width=4.3cm]{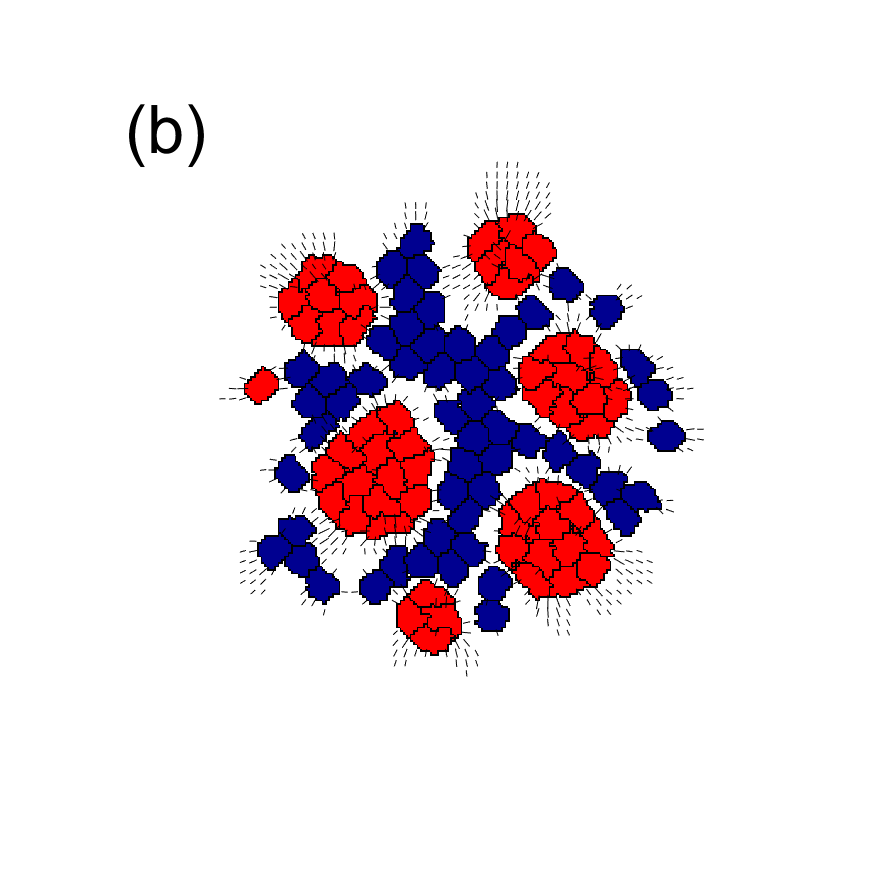}
\includegraphics[clip,width=4.3cm]{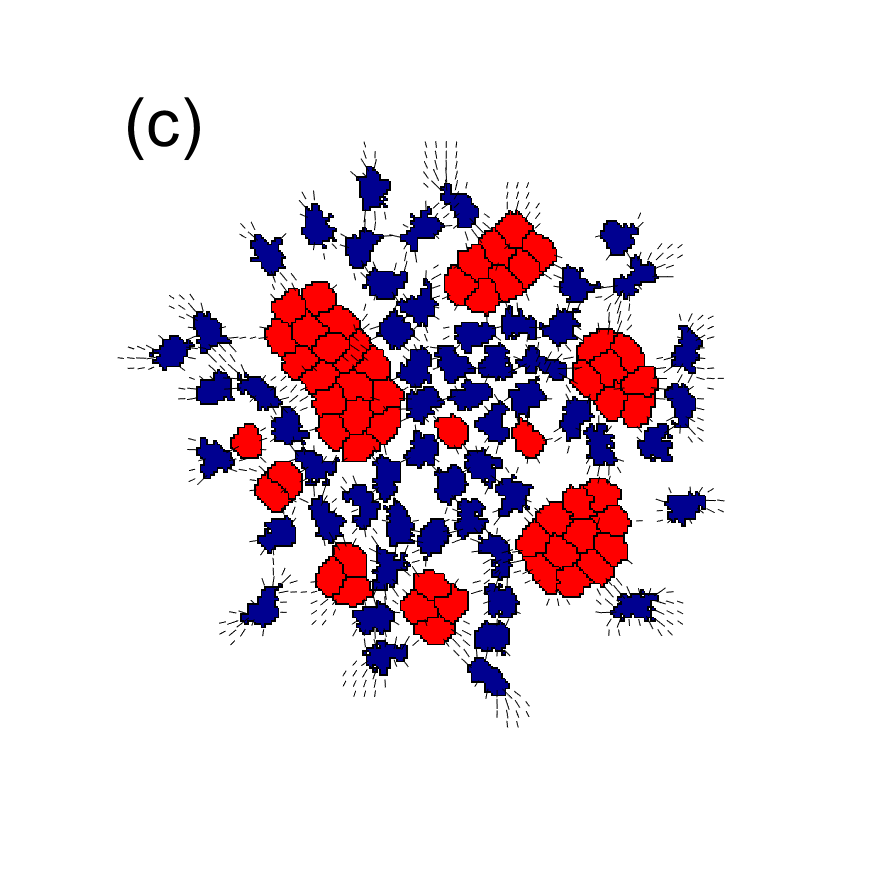}\\
\includegraphics[clip,width=4.3cm]{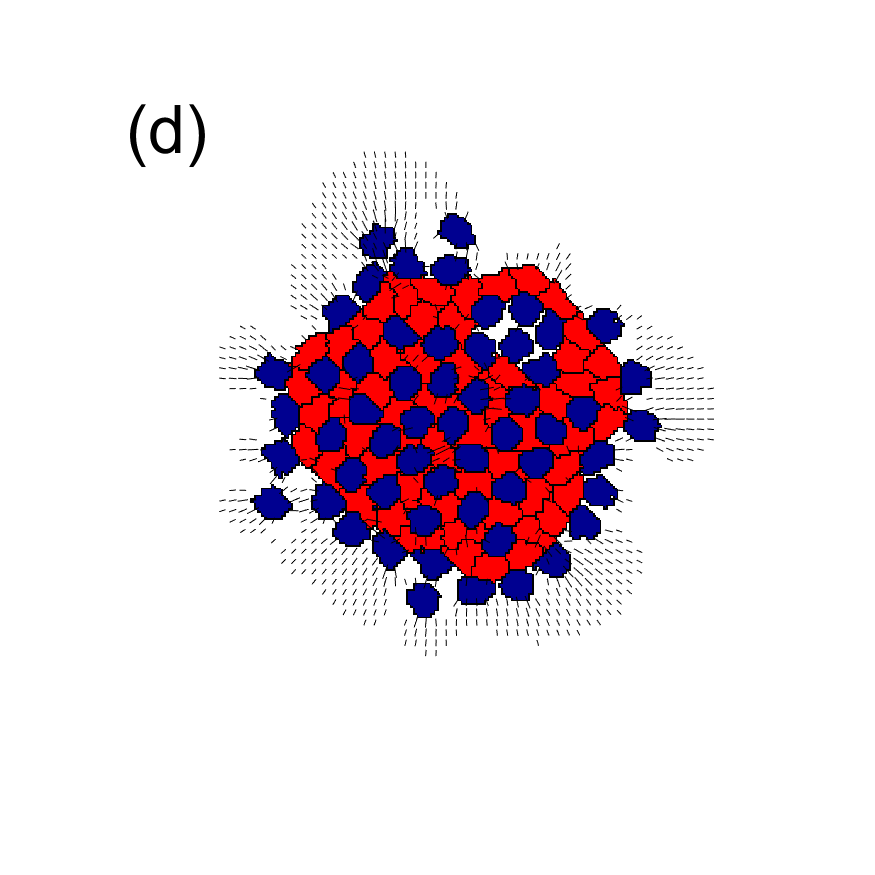}
\includegraphics[clip,width=4.3cm]{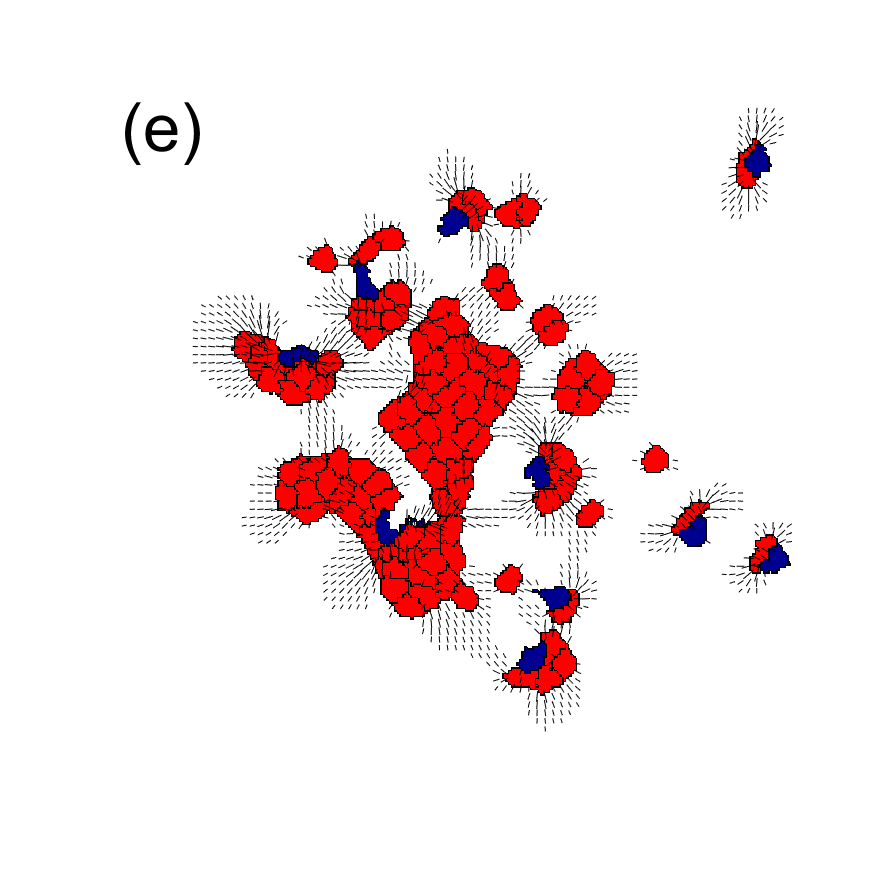}
\includegraphics[clip,width=4.3cm]{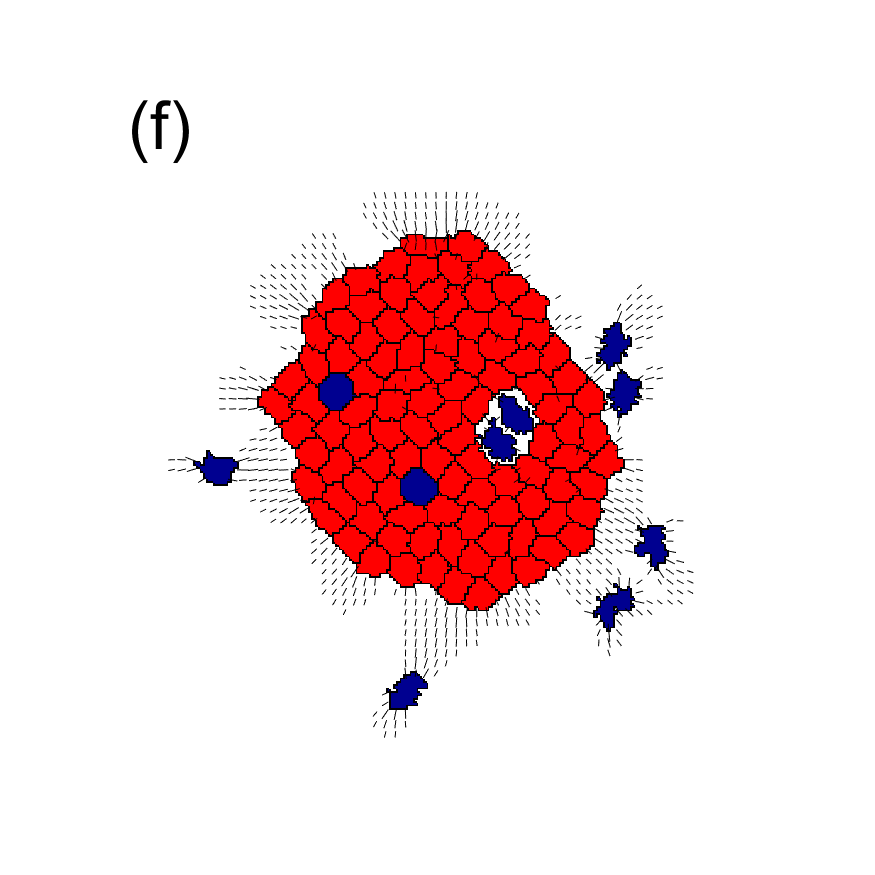}\\
\includegraphics[clip,width=4.3cm]{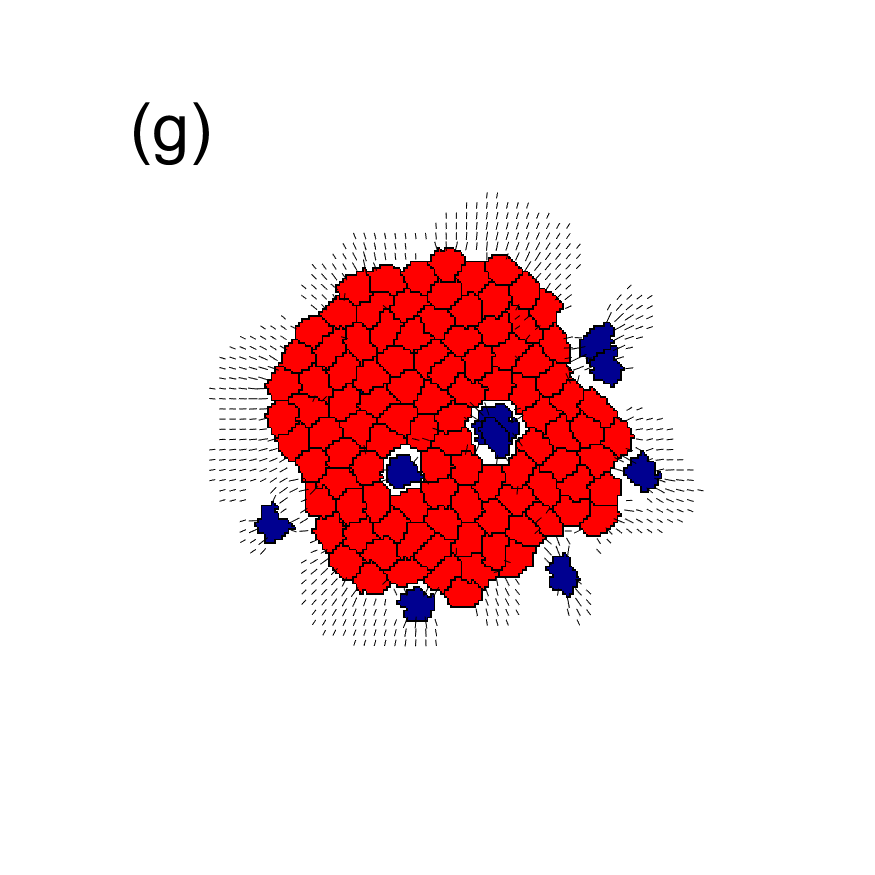}
\includegraphics[clip,width=4.3cm]{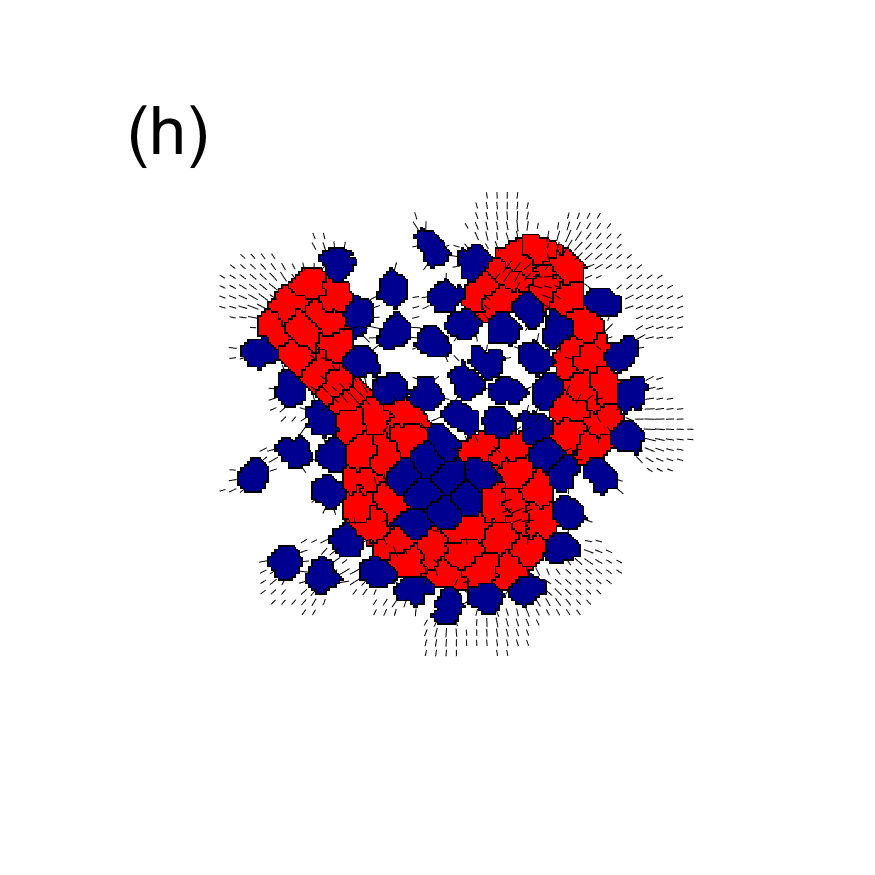}
\includegraphics[clip,width=4.3cm]{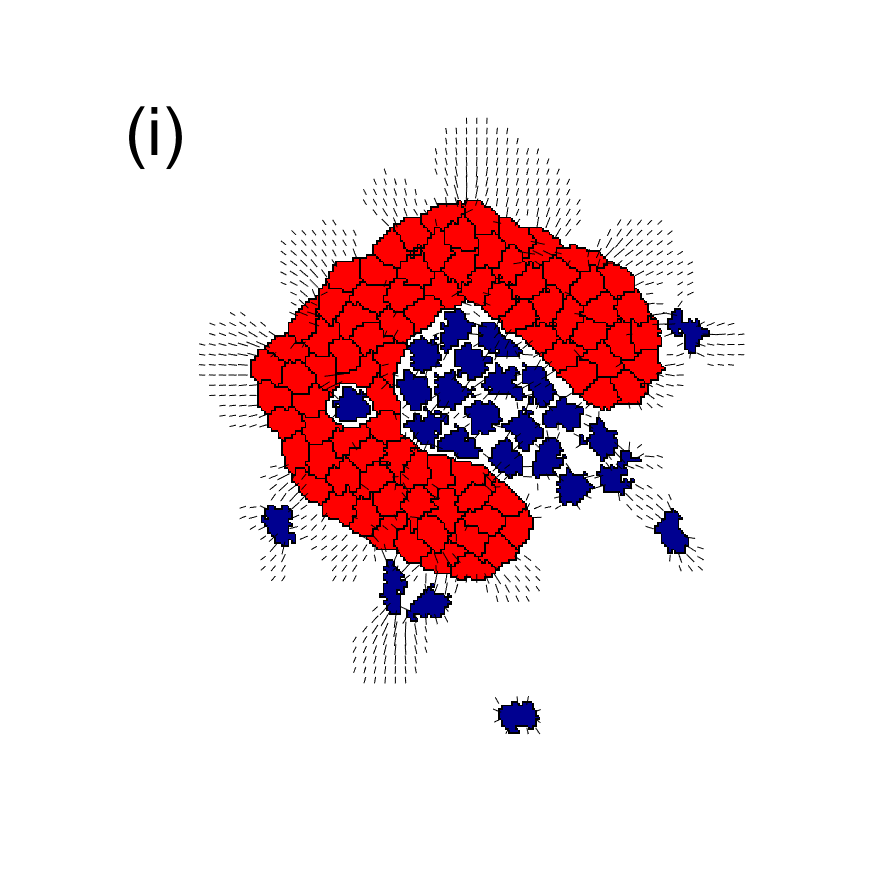}
\caption{{\bf Changes in adhesion parameters produce different patterns.}
A total of $N_{total}=111$ cells are randomly placed inside a circular enclosure, with $N_m$ motile  cells (blue) and passive adhesive cells (red) having $j_{cs}=1$ and $E_{\theta c}=E_{\theta m}=1$ kPa in Panels (a), (b), (d), (h), and $E_{\theta c}=1$ kPa, $E_{\theta m}=15$ kPa in Panels (c), (e), (f), (g), (i). Relative adhesion parameters are ($J_{kl} = j_{kl} J_{ref}$): (a) $j_{cc}=1/2$, $j_{cm}=2$, $j_{ms}=1$, $j_{mm}=2$, $N_m=56$. (b) same as (a) except $j_{cm}=11$. (c) $j_{cc}=1$, $j_{cm}=4$, $j_{ms}=1/4$, $j_{mm}=6$, $N_m=56$. (d) same as (a) except $j_{mm}=10$. (e) $j_{cc}=1/2$, $j_{cm}=2$, $j_{ms}=1$, $j_{mm}=12$, $N_m=10$. (f) $j_{cc}=j_{cs}=1$, $j_{cm}=4$, $j_{ms}=1/4$, $j_{mm}=6$, $N_m=10$. (g) same as (f) except  $j_{cm}=10$, $j_{ms}=1/2$, $j_{mm}=1/2$. (h) same as (a) except  $j_{mm}= 3$. (i) same as (e) except $j_{cm}=10$, $j_{ms}=1/4$. At $t=3000$ MCTS, the corresponding pattern types are: (a) 1, (b) 2, (c) 3, (d) 4, (e) 5, (f) 6, (g) 7, (h) 1,2,3, and (i) 3,6.}
\label{fig2}
\end{figure}

We consider two cell types (passive and motile). Their adhesion dynamics depend on their YMT  ($E_{\theta c}$ for passive, $E_{\theta m}$ for motile cells), the cell populations ($N_m$ motile, $N_c$ passive cells with $N_{\text{total}}=N_c+N_m$), and the cell-cell and cell-substrate adhesion parameters, $J(\sigma(\mathbf{x}), \sigma(\mathbf{x}'))$, in Eq~(\ref{eq5}). The latter adopt different values with respect to a reference adhesion $J_\text{ref}$, which we denote by
\begin{itemize}
    \item $J_{cs}= j_{cs} J_{\text{ref}}$: passive cell-substrate adhesion,
    \item $J_{cc}= j_{cc} J_{\text{ref}}$: passive cell-passive cell adhesion,
    \item $J_{cm}= j_{cm} J_{\text{ref}}$: passive cell-motile cell adhesion,
    \item $J_{mm}= j_{mm} J_{\text{ref}}$: motile cell-motile cell adhesion,
    \item $J_{ms}= j_{ms} J_{\text{ref}}$: motile cell-substrate adhesion.
\end{itemize}
See Fig.~\ref{fig1}. Lower (resp.\ higher) adhesion parameter values facilitate (resp.\ impede) adhesion. The relative values between adhesion parameters are critical: If $j_{cs} = j_{cm}> j_{cc}$ for example, passive cells prefer to adhere to each other preferably than adhering to the substrate or to motile cells. Through a qualitative analysis, we have identified in Appendix \ref{ap:a} seven dynamic patterns based on variations in adhesion parameters, substrate stiffness, and cell populations:

\begin{enumerate}
    \item \textit{Clusters of passive and motile cells sharing boundaries, each cell preferring their own type:} When $j_{cc} < j_{mm} \approx j_{cm} \approx j_{cs} \approx j_{ms}$, cells prefer clustering with their own type, and secondarily with the opposite type, over sticking to the substrate. See Fig.~\ref{fig2}(a).
    
    \item \textit{Separate clusters consisting of only passive or motile cells with no inter-species adhesion:} Strong intra-species adhesion ($j_{qq}$, $q = c, m$) and weak inter-species adhesion, $j_{cm} \gg j_{qq}$, promote segregation. See Fig.~\ref{fig2}(b).

    \item \textit{Separated passive and motile cells, some single motile cells:} Motile cells exhibit weak interactions, while passive cells cluster tightly. Flexible substrates enhance freedom of motile cells. For structured separations, this pattern requires $j_{cm}, j_{mm} \gg j_{ms}$ and balanced populations, $N_m \approx N_c$. See Fig.~\ref{fig2}(c).

    \item \textit{Disordered compact pattern with motile cells forming the outer rim:}  Cells of different phenotype cluster together when inter-phenoype adhesion dominates, $j_{cm} \ll j_{mm}$. See Fig.~\ref{fig2}(d).

    \item \textit{Motile cells push passive cells apart yielding a fragmented pattern:} Motile cells exert pressure, disrupting cohesion of passive cells. Differences in stiffness thresholds, $E_{\theta c} \neq E_{\theta m}$, and $j_{ms}< j_{cm}<j_{mm}$ drive this pattern. See Fig.~\ref{fig2}(e).

    \item \textit{Single motile cells escape from a cluster of passive cells, motile cells are at the periphery and moving away:}  Motile cells surround passive cell groups when $j_{ms} \ll j_{mm}, j_{cm}$ and $N_m \ll N_c$. Passive cells act as the central reference group. See Fig.~\ref{fig2}(f).

    \item \textit{Cluster metastasis: Motile cells form clusters around passive cells:} Motile cells group together in a `metastasis' configuration around passive cells when $j_{mm} \leq j_{ms} \ll j_{cm}$. See Fig.~\ref{fig2}(g).
\end{enumerate}

Figure~\ref{fig2} illustrates these patterns. Figs.~\ref{fig2}(a), \ref{fig2}(b), \ref{fig2}(d), and \ref{fig2}(h) have equal YMTs in Eq.~\eqref{eq9}, $E_{\theta c}=E_{\theta m}=1$kPa, whereas Figs.~\ref{fig2}(c), \ref{fig2}(e), \ref{fig2}(f), and \ref{fig2}(g) have different YMT, $E_{\theta c}=1$ kPa, $E_{\theta m}=15$ kPa. Only cells with different YMT show motile cells able to travel, thereby describing possible metastatic scenarios. Figure~\ref{fig2} shows cellular configurations of the main patterns 1 to 7 as well as  mixtures of different patterns in Figs.~\ref{fig2}(h) and \ref{fig2}(i). To ensure reproducibility of our results, Appendix \ref{ap:a} enumerates 21 configurations and lists in a table the conditions under which different patterns are found. To understand the table, take for example entry \# 24 that combines configurations 2, 5, 7, 9, 21. It corresponds to an aggregate of 111 cells, of which 10 (a much smaller number) are motile (configurations 2,5) having different YMTs as in configuration 7. Thus, motile cells advance more easily than passive ones. The relative adhesion parameters are $j_{ms}=1/4$, $j_{cs}=j_{cc}=j_{mm}=1$, $j_{cm}=10$, which correspond to configurations 9, 21. The adhesion parameter between motile cells and substrate is smaller than that between passive cells and substrate and motile-motile and passive-passive cells, whereas the adhesion parameter between motile and passive cells is much higher. With these values, motile cells tend to disassociate themselves from passive cells whereas adhesions between like cell types or with the substrate are much more favorable. The resulting pattern is number 6 corresponding to Fig.~\ref{fig2}(f). Mixtures of patterns occur with intermediate parameter values, reflecting locally dependent choices by cells; see Appendix \ref{ap:a}. 

\section{Active migration forces}\label{sec:4}
To achieve directed migration of the motile cells, we introduce a global long range active migration force $\mathbf{f}_m$. Long range force transmission of mechanical origin is responsible for phenomena such as waves in antagonistic migration assays \cite{rod17}, stress fibres mechanics \cite{oak17}, collective cell durotaxis \cite{sun16}, etc. The migration force should be incorporated into the finite element part of the model by modifying the Navier equation:
\begin{subequations}\label{eq10}
\begin{eqnarray}
\mathbf{K}\mathbf{u}=\mathbf{f}+\mathbf{f}_m. \label{eq10a}
\end{eqnarray}
Here, we propose an idealized simple long range migration force: $\mathbf{f}_m$ is zero for passive cells, whereas it depends on the distance to a point of reference (push force) or to a point of attraction (pull force) for motile cells. A push migration force on node $i$ of the motile cells is
\begin{eqnarray}
\mathbf{f}_{mi}=\mu \mathbf{d}_{mi} = \mu(\mathbf{d}_{m}-\mathbf{d}_{i}), \label{eq10b}
\end{eqnarray}
where $\mathbf{d}_m$ is the position of a reference point opposite to the target location, and $\mathbf{d}_i$ is the position of node $i$ of the motile cell. We have kept the same tension per unit length as in Eq.~\eqref{eq1}, which is calibrated to the observed values of the total cell traction \cite{vanOers}. Raising or lowering the value of $\mu$ in Eq.~\eqref{eq10b} with respect to that in Table~\ref{t1} does not change the outcome of the numerical simulations (see below). The force $\mathbf{f}_m$ points towards $d_m$, away from the target point, and modifies the displacement $\mathbf{u}$, whose discrete gradient for an extension towards the target point is $\mathbf{0}-\mathbf{f}_{mi}$. In turn, this generates strains $\mathbf{\varepsilon}$ in Eq~(\ref{eq7}) whose magnitude increases with $|\mathbf{d}_{m}-\mathbf{d}_{i}|$. These strains enter the CPM through the durotaxis Eq.~\eqref{eq6} and push the cell towards the target location (under cellular extension towards it). As an alternative to Eq.~\eqref{eq10b}, the pull migration force is
\begin{eqnarray}
\mathbf{f}_{mi}=\mu \mathbf{d}_{ia} = \mu(\mathbf{d}_{i}-\mathbf{d}_{a}),  \label{eq10c}
\end{eqnarray}
\end{subequations}
where $\mathbf{d}_a$ is the position of the target location (attraction point). The force $\mathbf{f}_{mi}$ of Eq.~\eqref{eq10c} also points away from the target location $\mathbf{d}_a$ and pulls the cell towards the attraction point. The difference between push and pull forces is that the push force is stronger the further the distance to $\mathbf{d}_m$ is (and therefore the shorter the distance to $\mathbf{d}_a$ is), whereas the pull force is weaker the shorter the distance to $\mathbf{d}_a$ is. The push force provides a more efficient attraction to the target point at the cost of a greater dispersion of motile cells. From now on, {\em we use the push force as our migration force of choice}.

\begin{figure}[!h]
\includegraphics[clip,width=4.3cm]{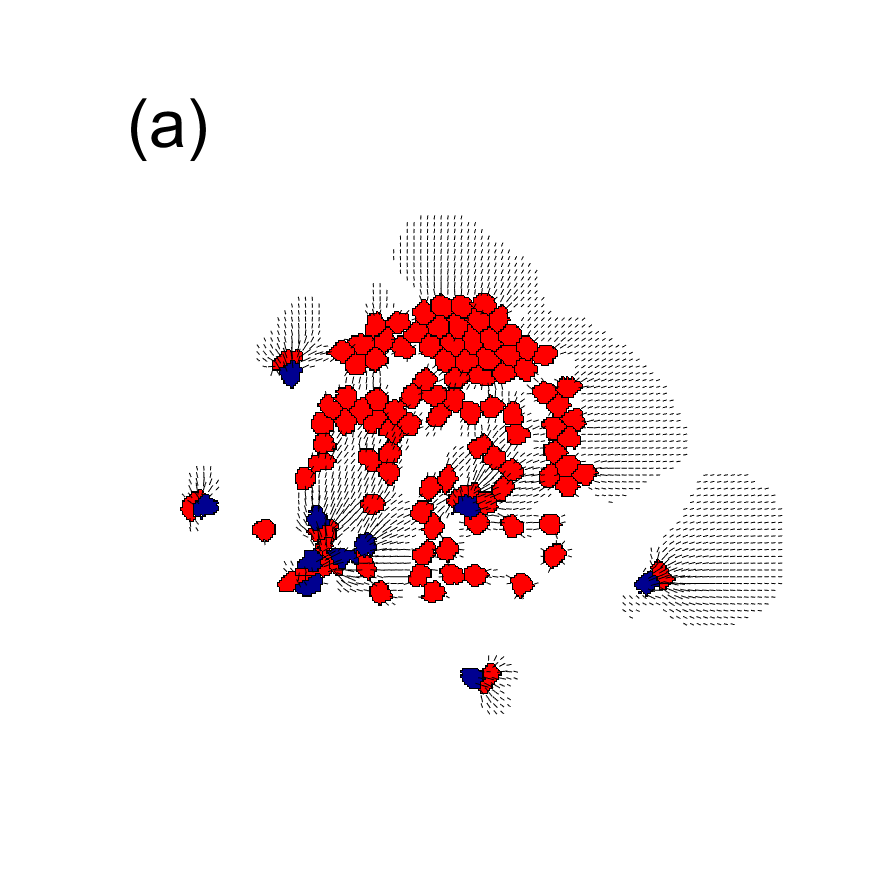}
\includegraphics[clip,width=4.3cm]{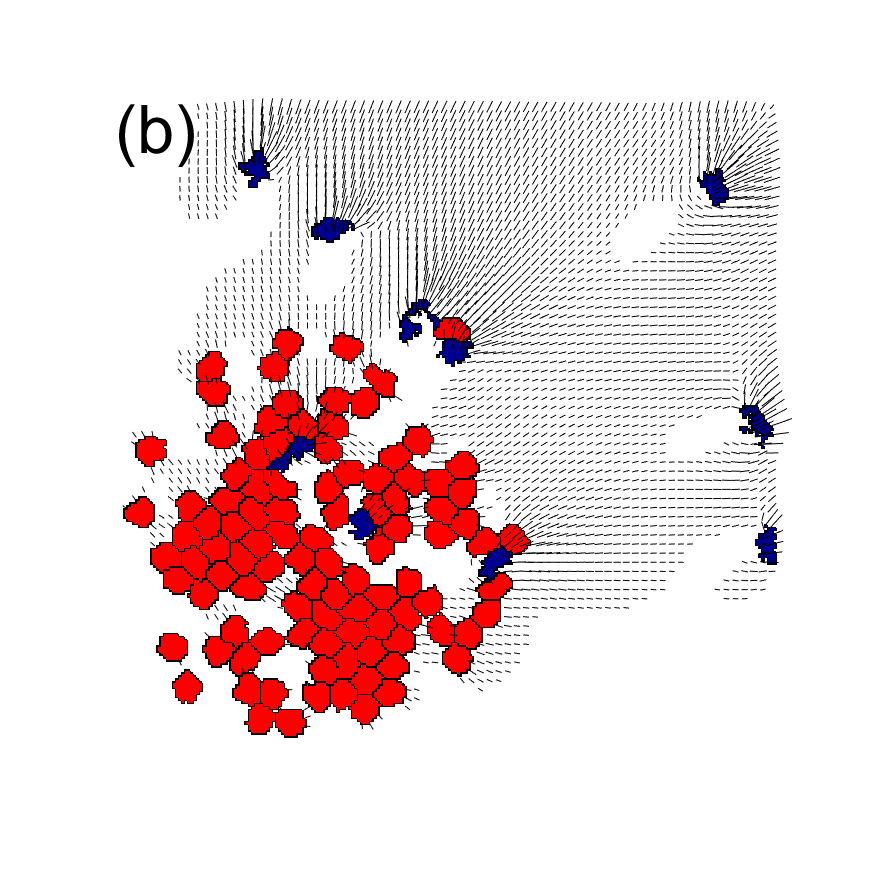}
\includegraphics[clip,width=4.3cm]{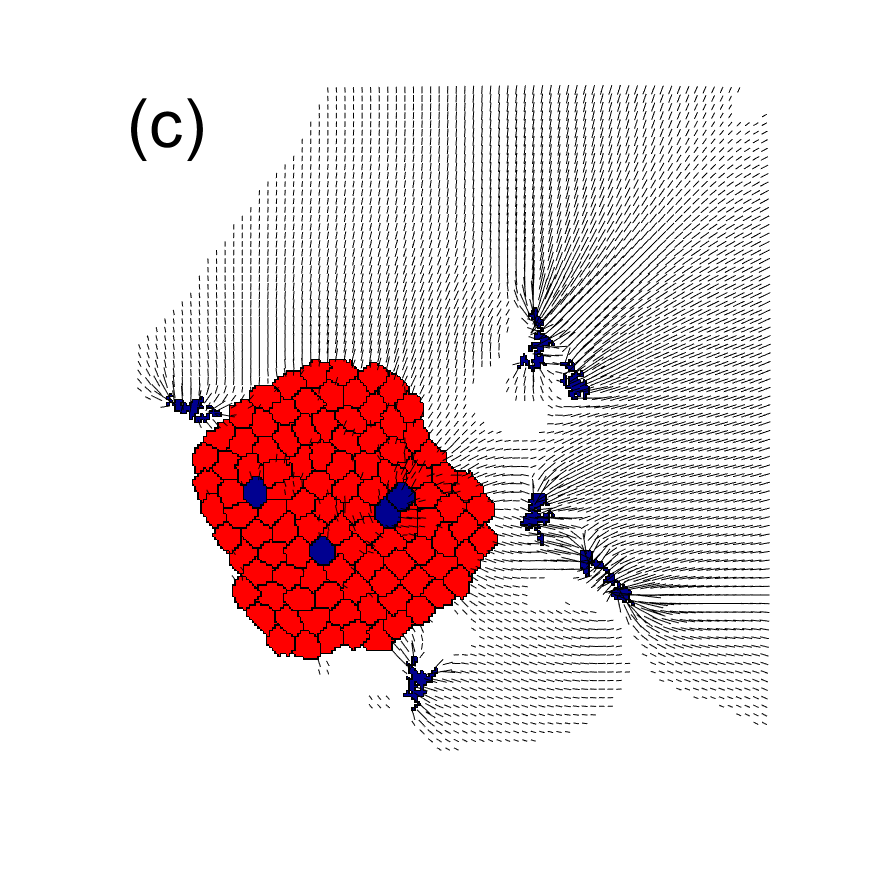}
\caption{{\bf Simulated migration force towards an attraction point} on the upper right corner.
Migration of motile blue cells ($E_{\theta m}= 15$ kPa) from passive red cell aggregate ($E_{\theta c}=1$ kPa), $j_{mm} = j_{cm} =j_{cc}=2$, $j_{ms}=j_{cs}=1$. Substrate strains are shown by black arrows. With Fig~\ref{fig1}(a) as initial condition, result at 1500 MCTS of: (a) choice A, (b) choice B, (c) choice B with $j_{cc} = j_{cs} = 1$, $j_{mm} = 6$, $j_{cm} = 4$, $j_{ms} = 1/4$. See Video S2 \cite{suppl}.}
\label{fig3}
\end{figure}

\subsubsection{Performance of the migration force}
Incorporating the active push migration force to the CPM means solving Eqs.~\eqref{eq10} and then implementing the Metropolis algorithm according to Eqs.~\eqref{eq2}-\eqref{eq9}. The strain gradient created by the migration force directs  motile cells towards the point of attraction and away from the reference point. With respect to the YMT, we have two choices:
\begin{itemize}
\item[A.] In extension (resp.\ retraction), adopt the YMT of the target (resp.\ source) pixel. This extends the extension-retraction symmetry of durotaxis under traction forces $\mathbf{f}$ to the combined force $\mathbf{f}+\mathbf{f}_m$.
\item[B.] Always adopt the YMT of the target pixel for $\mathbf{f}+\mathbf{f}_m$.
\end{itemize} 
Figure~\ref{fig3} shows how motile blue cells respond to the migration force by advancing towards the attraction center whereas the passive red cells are insensitive to it. The black arrows represent the strains in the substrate, which are influenced by the migration force. Choice A does not result in a successful migration of motile cells to the attraction center, cf Fig.~\ref{fig3}(a). Choice B  leads to migration if min$(j_{mm}, j_{ms}) \geq 1$, cf Fig.~\ref{fig3}(b). If the latter condition does not hold, the motile cells may not progress to the attraction center in a regular shape, cf Fig.~\ref{fig3}(c). 
Migration becomes negligible for some configurations of adhesion parameters when the Hamiltonian variation due to stiffness is greater than that of migration. Choice B may produce motile cells of irregular shape, cf Figs.~\ref{fig3}(b) and\ref{fig3}(c). Furthermore, adding biochemical energy changes to $H$ requires calibrating the adhesion parameters in choice B to attain collective invasion. Thus, the question is: How could we make migration more efficient?

 \subsubsection{Fractional step CPM}
To enhance motility, we propose a fractional time step method: Displacements due to cell traction forces $\mathbf{f}$ are computed at time $t=n$ ($n=0,1,2, \ldots$) and Eqs.~\eqref{eq1}-\eqref{eq9} solved. Then we find a `migration' displacement $\mathbf{u}_m$ from solving 
\begin{subequations} \label{eq11}
\begin{eqnarray}
\mathbf{K}\mathbf{u}_m = \mathbf{f}_m,  \label{eq11a}
\end{eqnarray}
at the intermediate time step $t=n+1/2$. $\mathbf{u}_m$ produces a strain
\begin{eqnarray}
\mathbf{\varepsilon}_m = \mathbf{B}\mathbf{u}_{em},\label{eq11b}
\end{eqnarray}
from which the Hamiltonian variation is
\begin{eqnarray}
\Delta H_\text{mig}\! = -g(\mathbf{x},\mathbf{x}')\lambda_{mig}\!\left[h(E(\epsilon_{m1}))(\mathbf{v}_1\cdot\mathbf{v}_m)^2\! + h(E(\epsilon_{m2}))(\mathbf{v}_2\cdot\mathbf{v}_m)^2\right]\!\!.\quad \label{eq11c}
\end{eqnarray}
We give equal weight to migration and stiffness-induced variations of energy, $\lambda_\text{mig} = \lambda_\text{duro}$. In Eq.~\eqref{eq11c}, the threshold energy $E_\theta$ is always that of the target pixel, both in extension and in retraction. This breaks the symmetry extension-retraction existing in the energy variation of Eq.~\eqref{eq6} and facilitates motion of the cells in a preferred direction.
\end{subequations}
The resulting new configuration is accepted according to the probability given by Eq.~\eqref{eq2} with energy $\Delta H=\Delta H_\text{volume}+\Delta H_\text{contact}+\Delta H_\text{mig}$ instead of Eq.~\eqref{eq3}.

The two approaches can be summarized as follows:
\begin{itemize}
    \item \textit{Single time step:} Combines traction and migration forces in a single equation ($\mathbf{K}\mathbf{u} = \mathbf{f} + \mathbf{f}_m$) at the same time step. The energy change is given by Eq.~\eqref{eq6} with the function $h(E)$ of Eq.~\eqref{eq9} in which the YMT may correspond to the target pixel in extension and it corresponds to the source pixel in retraction (choice A) or always to the target pixel (choice B). While simple to implement, this approach may not generate effective migration displacement as explained before.
    \item \textit{Fractional time step:} Separates traction ($\mathbf{K}\mathbf{u} = \mathbf{f}$) and migration ($\mathbf{K}\mathbf{u}_m = \mathbf{f}_m$) force computations into distinct time steps that have distinct strain tensors and therefore different energy changes. While the durotaxis (traction force) half step keeps the extension-retraction symmetry, in the migration half step, the YMT is that of the target pixel, both in extension and in retraction. The fractional step CPM allows iterative updates to the Hamiltonian for traction forces at $t=n$ and migration forces at $t=n+1/2$.
\end{itemize}
The flowcharts of the two approaches are presented in Appendix \ref{ap:b}.

\begin{figure}[!h]
\centering
\includegraphics[clip,width=3.2cm]{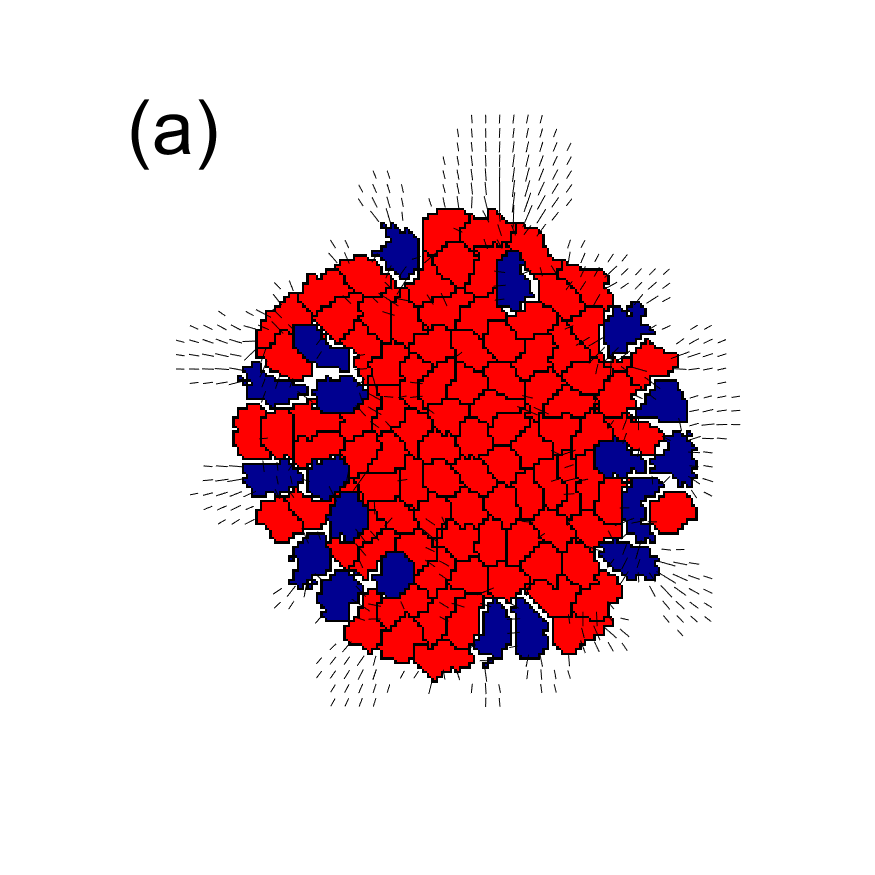}
\includegraphics[clip,width=3.2cm]{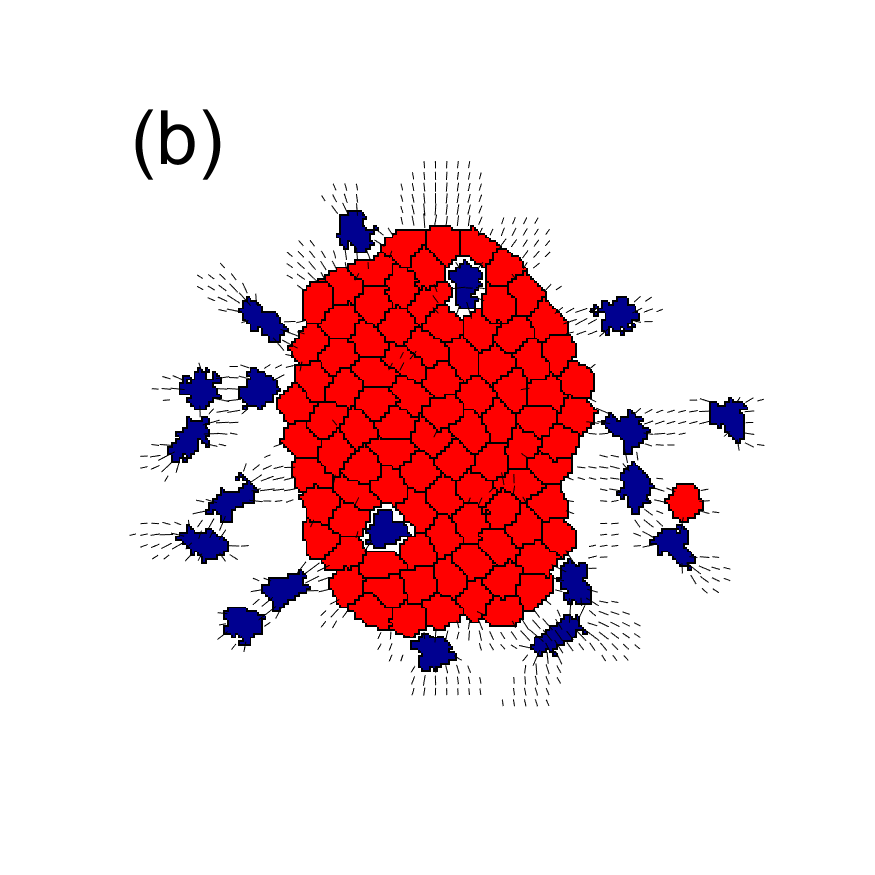}
\includegraphics[clip,width=3.2cm]{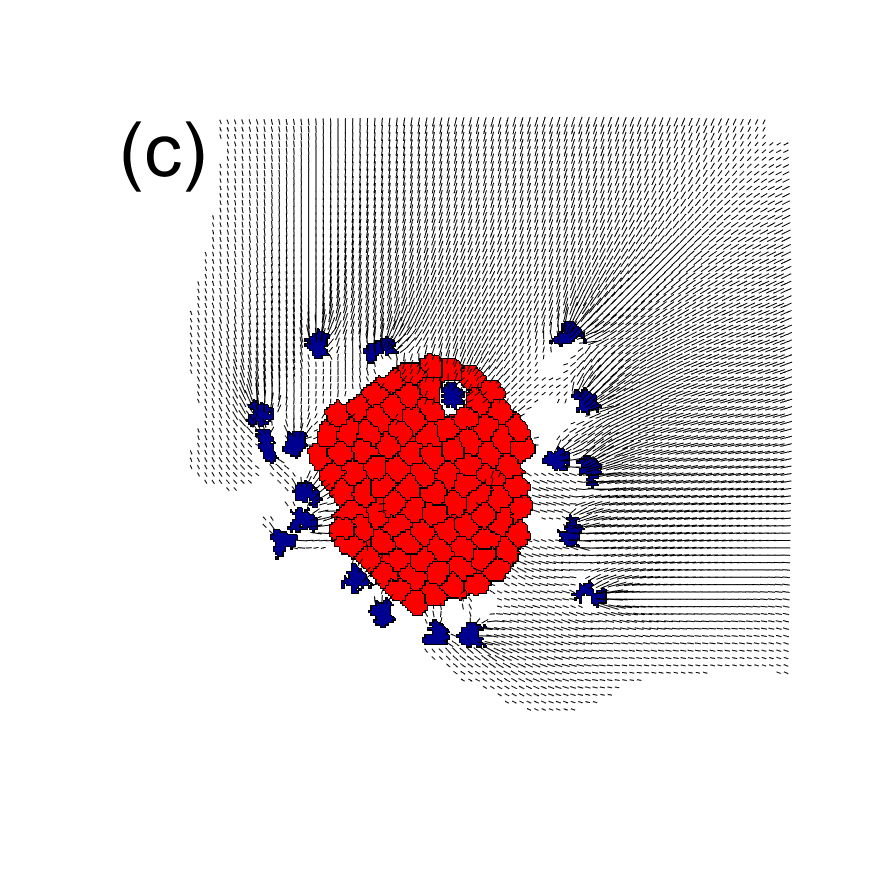}
\includegraphics[clip,width=3.2cm]{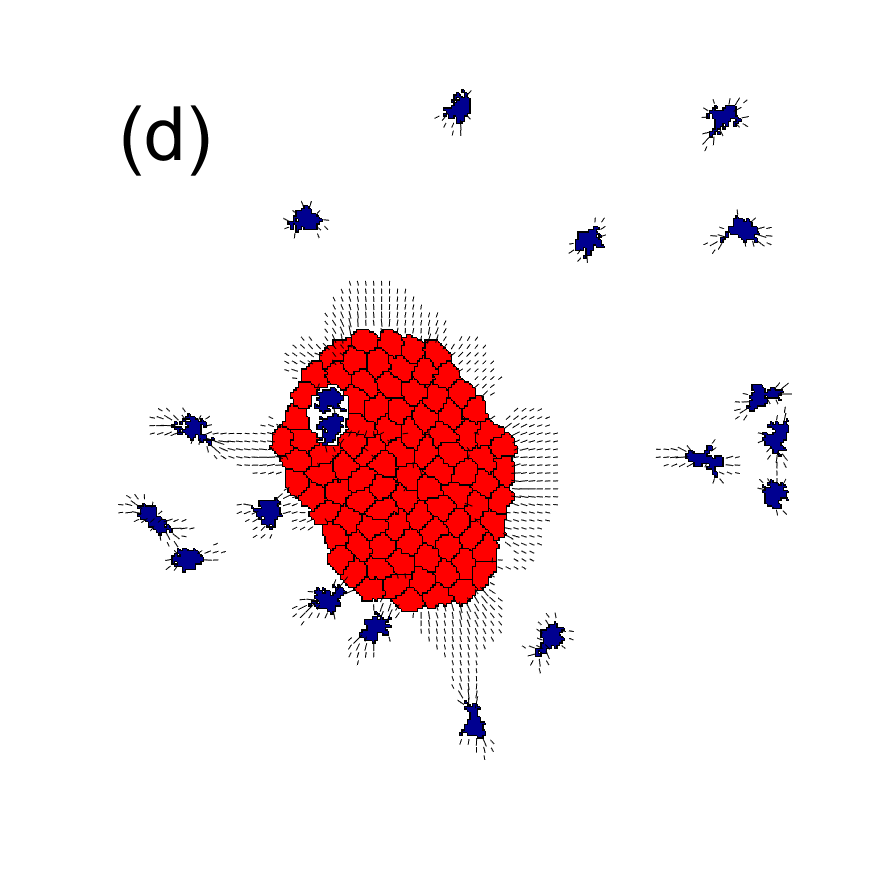}
\caption{{\bf Comparison of migration force implementations.} $N_{total}=111$ cells of which $N_m=19$ are motile (blue) are randomly placed inside a circular enclosure. Parameter configuration as in Fig.~\ref{fig2}(f): $E_{\theta c}=1$ kPa, $E_{\theta m}=15$ kPa, $j_{cs}=1$, $j_{cc}=1$, $j_{cm}=4$, $j_{ms}=1/4$, and $j_{mm}=6$. (a) Initial configuration. At $t=3000$ MCTS: (b) No migration force, (c) migration forces with the single time step approach, and (d) migration forces with the fractional time step approach. The attraction point is located on the top-right corner.}
\label{fig4}
\end{figure}

To recapitulate, as shown in Fig~\ref{fig4}, the fractional time step CPM ensures that motile cells reach the attraction point even under challenging adhesion configurations, outperforming the single time step approach. Consider Pattern 6 in Fig.~\ref{fig2}(f). Without the migration force, motile cells escape from the aggregate with no preferred direction; see Fig.~\ref{fig4}(b). What happens when we add reference/attraction points and a migration force as in Figs.~\ref{fig4}(c) and \ref{fig4}(d)? Cluster migration is easier to achieve as neighboring pixels with different YMTs generate a gradient directing movement towards the attraction point. Sometimes migration does not occur with traction and migration forces combined in a single time step, cf Fig.~\ref{fig4}(c), but the fractional step approach succeeds, cf Fig.~\ref{fig4}(d). 

 \begin{figure}[!h]
\centering
\includegraphics[clip,width=11.5cm]{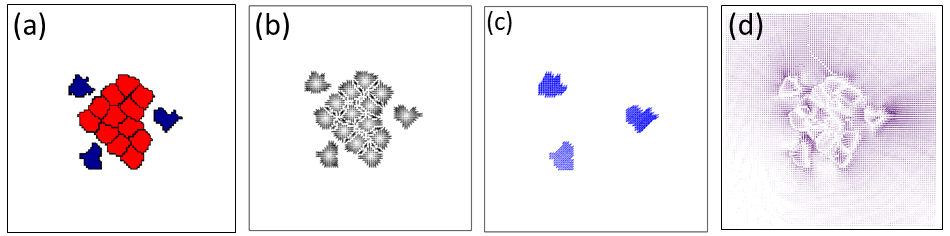}
\caption{ {\bf Forces and strains from the single step method.} A cluster of 12 E cells (red) and 3 motile M cells (blue) on its boundary has evolved 500 MCTS by the single step method: (a) final state, and density plots of: (b) traction force $f$, (c) push migration force $f_m$, (d) strain on ECM associated to force $f+f_m$. Parameter values are as in Fig.~\ref{fig2}(f) and the attraction point (not shown) is on the upper right corner. The initial configuration is similar to that of Fig.~\ref{fig4}(a).}
\label{fig5}
\end{figure}

\begin{figure}[!h]
\centering
\includegraphics[clip,width=9.5cm]{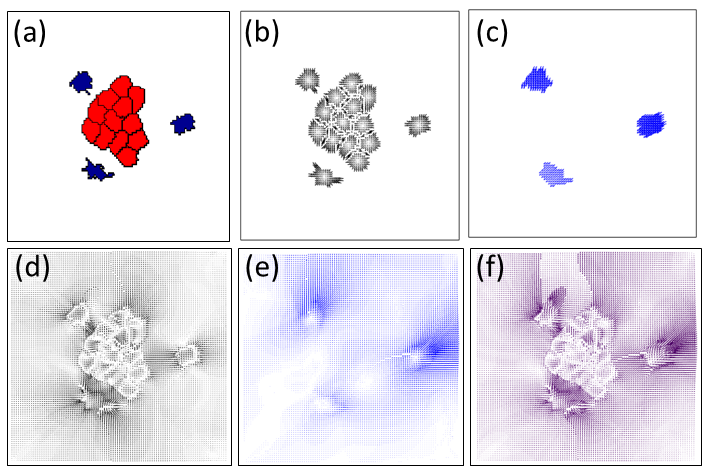}
\caption{ {\bf Forces and strains produced by the fractional step method.} As in Fig.~\ref{fig5}, (a) final state, and density plots of (b) stiffness force $f$, (c) migration push force $f_m$, (d) strain on ECM associated to stiffness force $f$, (e) same for the migration force $f_m$, (f) sum of strains  associated to stiffness and migration forces. The attraction point (not shown) is on the upper right corner.}
\label{fig6}
\end{figure}

Figures \ref{fig5} and \ref{fig6} show further details on forces and strains as generated by the single step and fractional step CPMs, respectively. Out of a 15-cell cluster, three M cells have left towards an attraction point on the upper right corner. Let us compare the strains in the ECM around the rightmost M cell (closer to the attraction point) as calculated by the single step CPM in Fig.~\ref{fig5}(d) and the same strains calculated by the fractional step CPM in Fig.~\ref{fig6}(f). The strains near the attraction point are greater when calculated by the fractional step CPM. 

\begin{figure}[!h]
\includegraphics[clip,width=4.3cm]{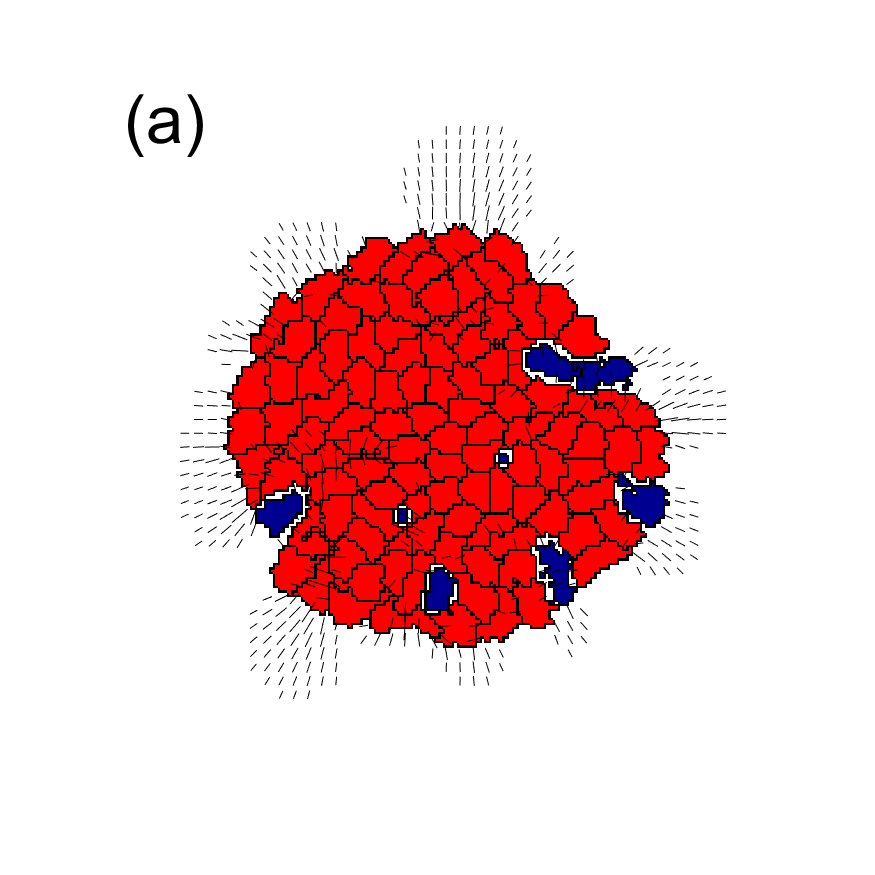}
\includegraphics[clip,width=4.3cm]{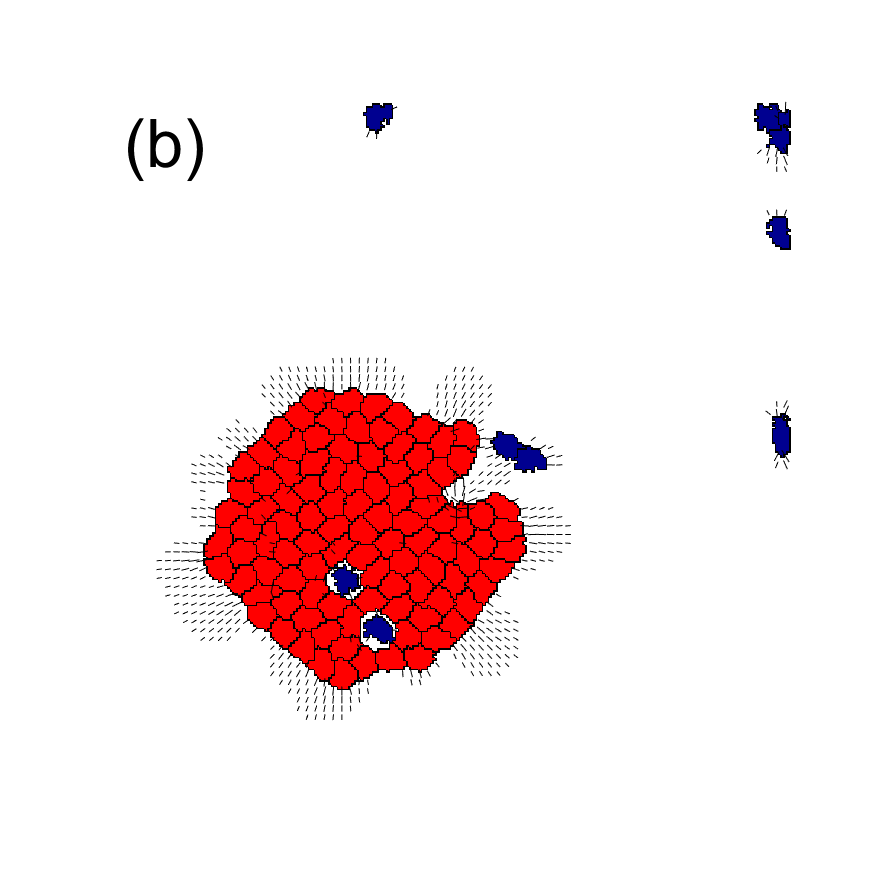}
\includegraphics[clip,width=4.3cm]{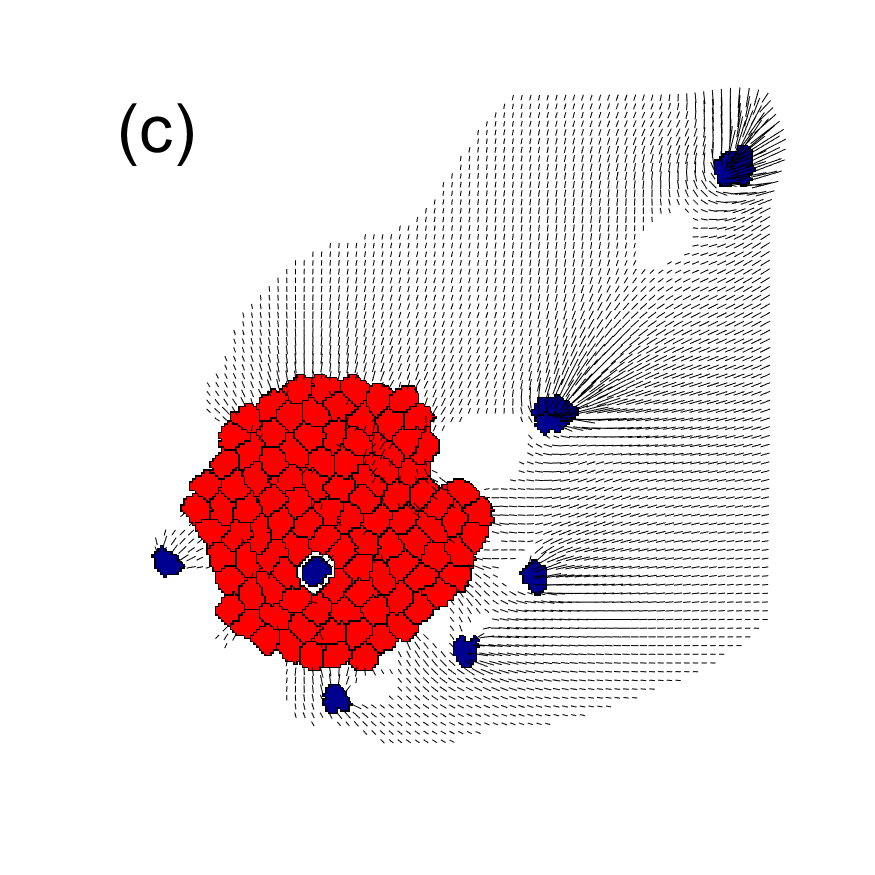}\\
\includegraphics[clip,width=4.3cm]{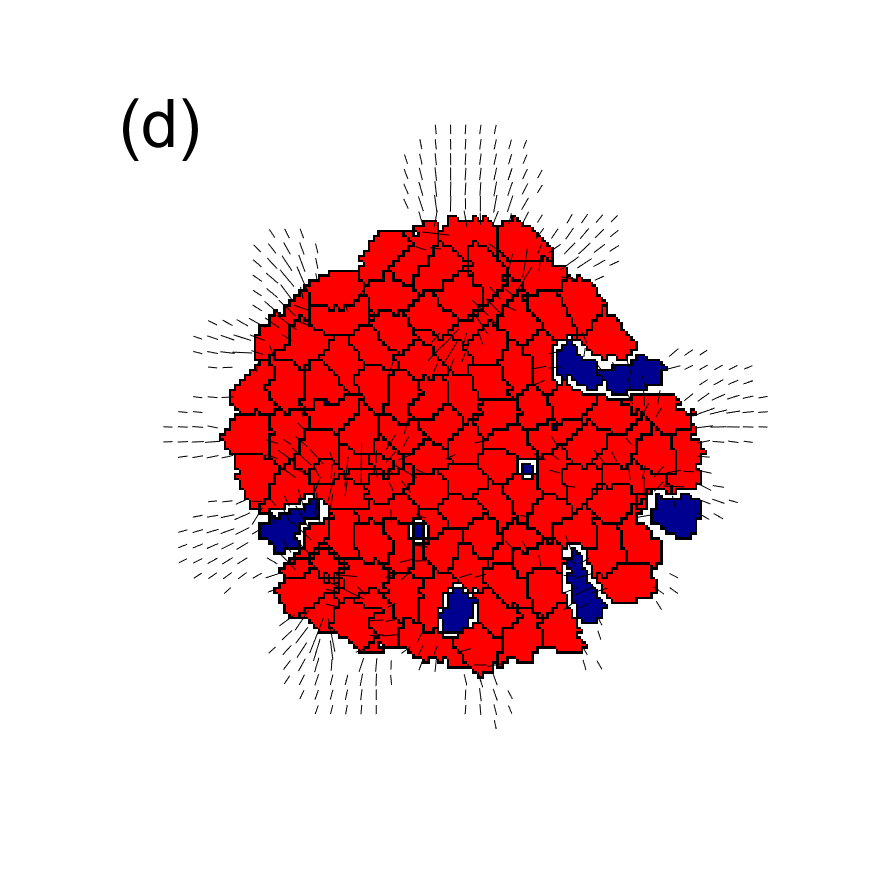}
\includegraphics[clip,width=4.3cm]{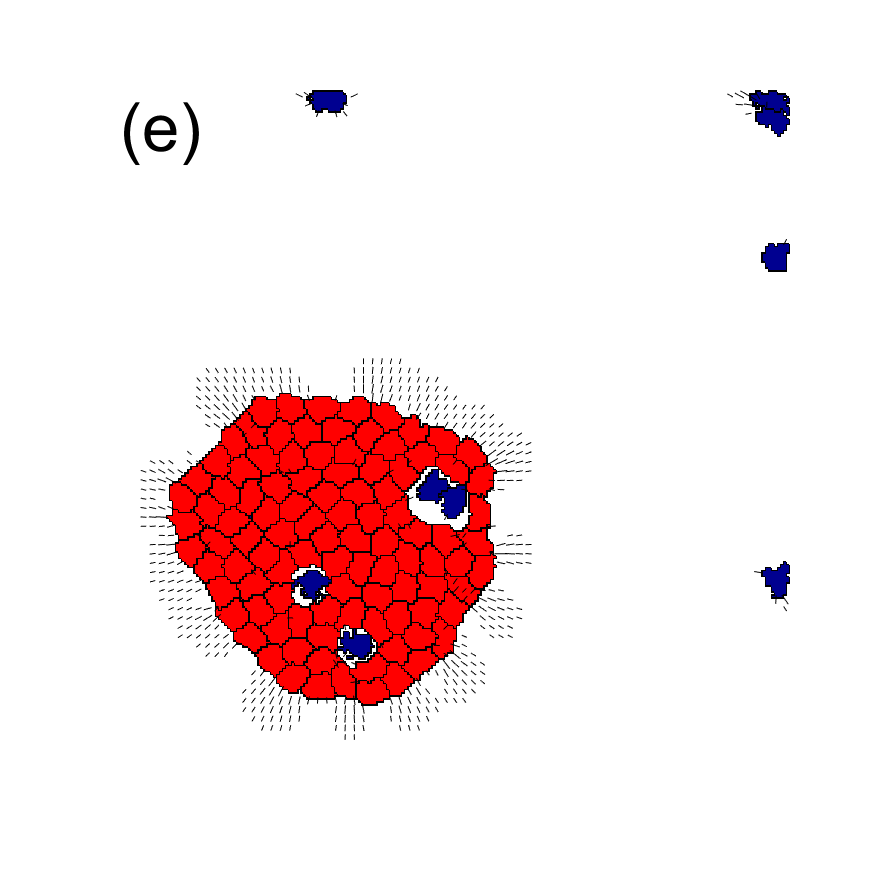}
\includegraphics[clip,width=4.3cm]{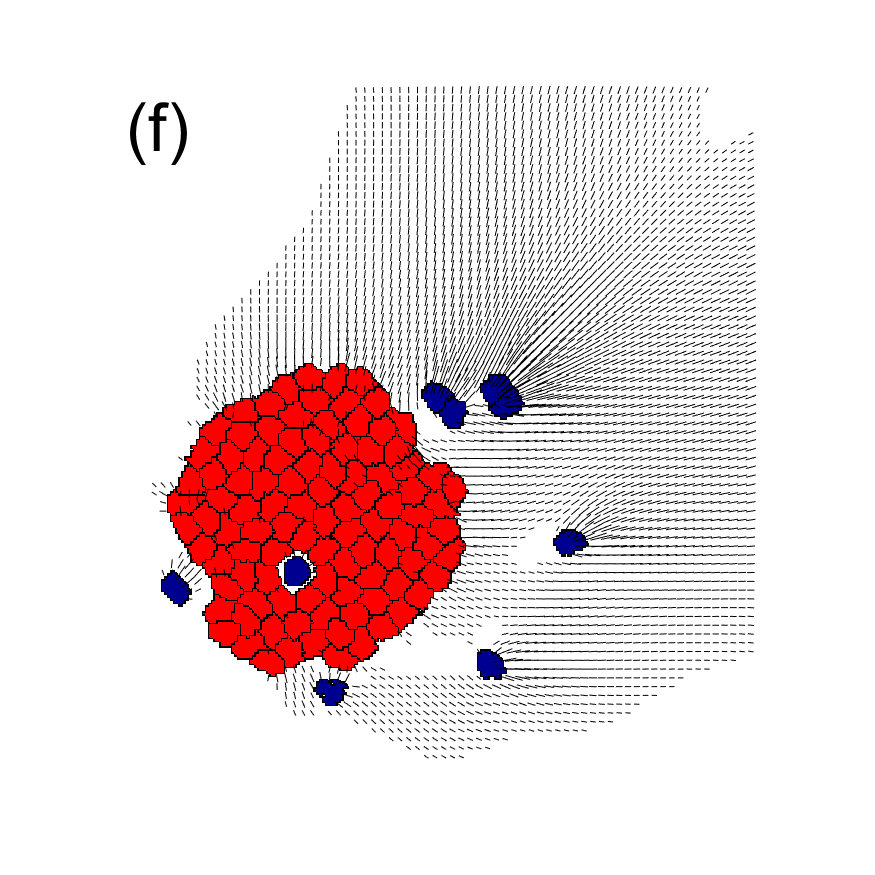}
\caption{{\bf Migration forces and invasion dynamics.} Initially, $N_m=10$ motile blue cells ($E_{\theta m} = 15$ kPa) and 101 passive red cells ($E_{\theta c} = 1$ kPa) form a circular aggregate with $j_{cs}=1$, $j_{cc}=1$, $j_{cm}=10$, and $j_{ms}=1$. First row has $j_{mm}=1/2$. Configurations at MCTS: (a) $t=15$, (b) $t=3000$ (fractional step), and (c) $t=3000$ (single step). Second row has $j_{mm}=1$. Configurations at MCTS: (d) $t=15$, (e) $t=3000$ (fractional step), and (f) $t=3000$ (single step). Attraction point (not shown) located in the upper-right corner of the enclosure. The full dynamics corresponding to panel (b) is in the Video S3 \cite{suppl}.}
\label{fig7}
\end{figure}

Figure~\ref{fig7} presents two simulations of migration of M cells from a cellular aggregate formed by E and M cells with other adhesion parameters that make it harder for E and M cells to separate. The single time step approach succeeds for M cells to reach the attraction point if $j_{mm}=1/2$ (low M-M cell adhesion, cf Fig.~\ref{fig7}(c)), but it does not produce migration if $j_{mm}=1$ (larger M-M cell adhesion, cf Fig.~\ref{fig7}(f)), whereas the fractional time step approach always enables migration; see Fig.~\ref{fig7}(b) and \ref{fig7}(e).
In qualitative terms, separating the steps of stiffness traction forces and migration forces is crucial. Stiffness forces as in Eq.~\eqref{eq1} have extension-retraction symmetry by choosing the YMT of the target pixel for extension and that of the source pixel for retraction. This ensures cohesive extension and retraction of the cells. The selection of theYMT of the target pixel both in extension and in retraction for the migration force of Eq.~\eqref{eq10b} enforces directional movement. Thus, the fractional CPM provides a robust framework for modeling active migration and stiffness-driven mechanics for different adhesion values and 
will be adopted in subsequent simulations.

\subsubsection{Computational Time Analysis.}  
Besides the qualitative differences in outcome explained above, the computational costs of the single step and fractional step methods vary significantly. We shall compare the results of the fractional step CPM in Figs.~\ref{fig4}(d), \ref{fig7}(b) and \ref{fig7}(e) with those of the single step CPM in Figs.~\ref{fig4}(c), \ref{fig7}(c) and \ref{fig7}(f). Times were measured using a HP Laptop 15s-fq5xxx with processor 12th Gen Intel$^\circledR$ Core\texttrademark\, i7-1255U (12 cores at 1.7 GHz), 16 GB RAM and 64 bit Windows 11 Home OS. 
\begin{itemize}
    \item Fig.~\ref{fig4}(d) (fractional step CPM) requires approximately 2.5 hours, compared to 45 minutes for Fig.~\ref{fig4}(c) (single step CPM).
    \item Fig.~\ref{fig7}(f) (single step CPM) is completed in 1 hour, while Fig.~\ref{fig7}(e) (fractional  step CPM) takes 2 hours and 15 minutes.
    \item For Fig.~\ref{fig7}(c) (single step CPM), the runtime is 1 hour and 45 minutes, compared to 2 hours and 15 minutes for Fig.~\ref{fig7}(b) (fractional step CPM).
\end{itemize}

When migration does not occur under the single step CPM, Fig.~\ref{fig7}(f), the fractional step CPM requires more than double the computational time (over 100\% longer, Fig.~\ref{fig7}(e)). However, in cases where migration from the aggregate succeeds under single time step, as in Fig.~\ref{fig7}(c), the additional computational cost for the fractional time step method of Fig.~\ref{fig7}(b) is modest, approximately 30 minutes, representing a 30\% increase ($t_f = 1.3 t_s$, where $t_f$ and $t_s$ denote runtimes for the fractional and single step CPMs, respectively). With respect to the single  step CPM, the computational cost of the fractional step CPM does not scale linearly with its increased complexity. Although the operation count doubles, the runtime only increases by a factor of 1.3. 
In conclusion, while the fractional step CPM introduces a slight additional computational cost, its robustness in enabling migration under varied adhesion parameters justifies its application in these scenarios.

\section{Results}\label{sec:5}

In this section, we illustrate invasion by single cells from a circular aggregate (tumor) and collective invasion. 

\begin{figure}[!h]
\centering
\includegraphics[clip,width=3.2cm]{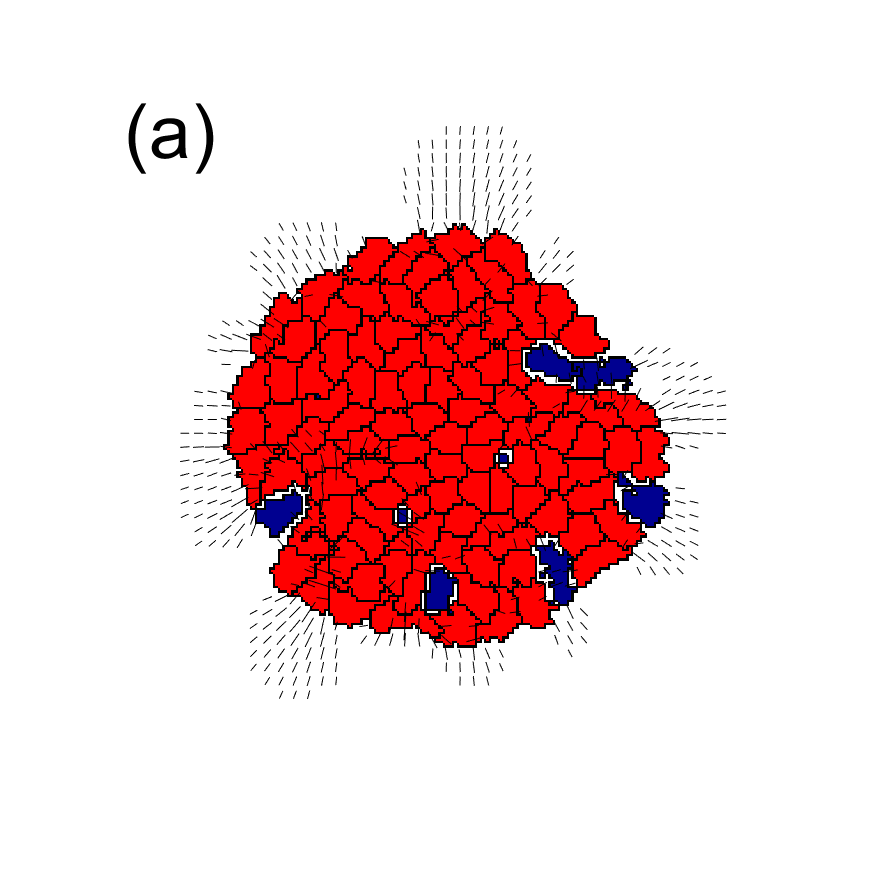}
\includegraphics[clip,width=3.2cm]{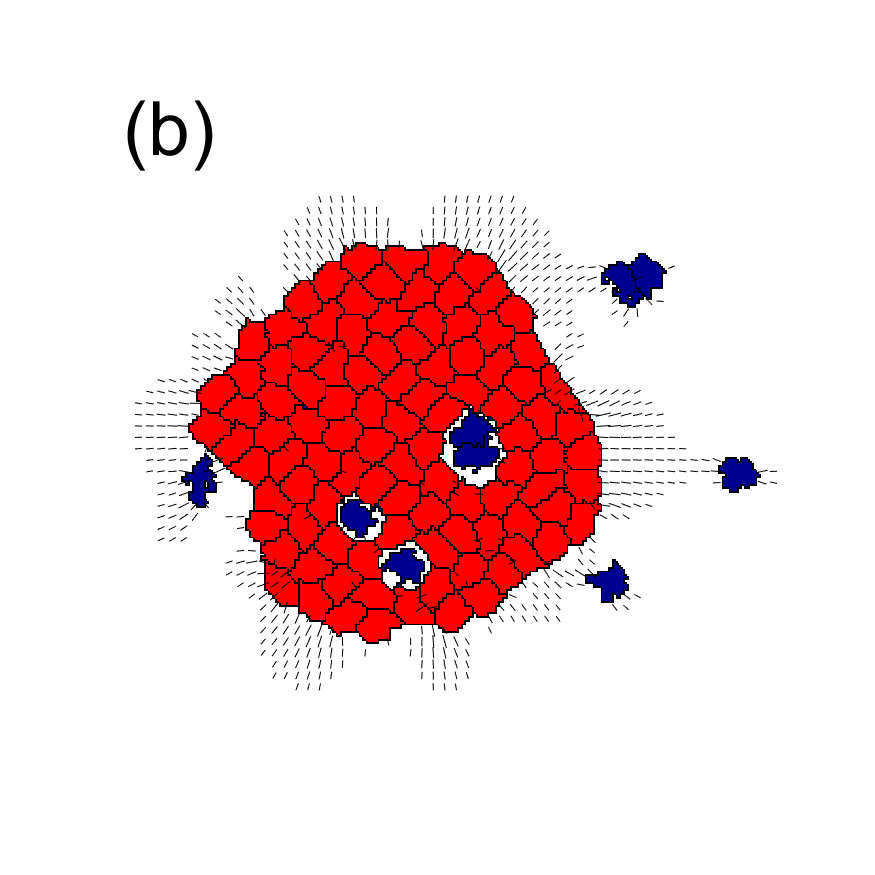}
\includegraphics[clip,width=3.2cm]{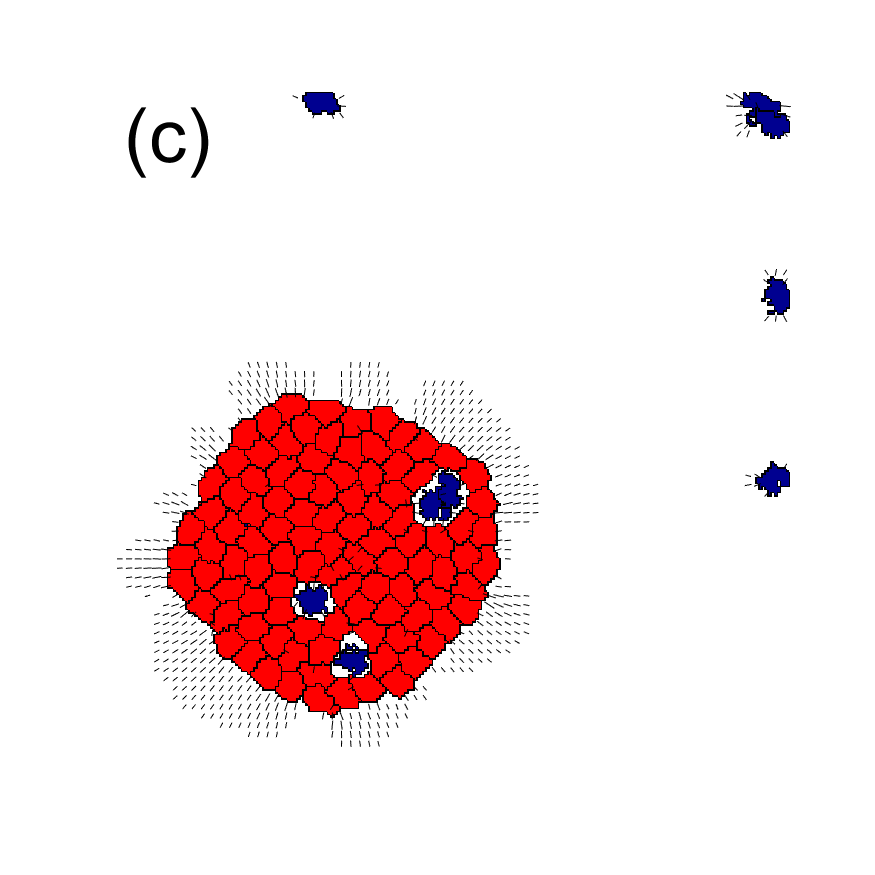}
\includegraphics[clip,width=3.2cm]{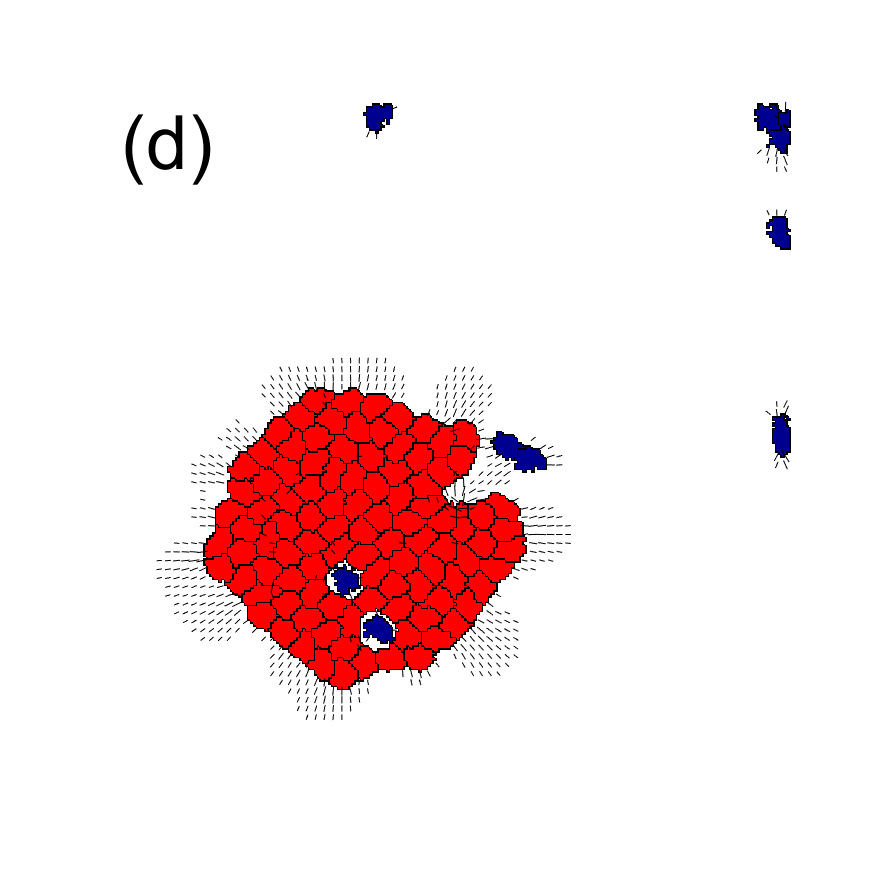}
\caption{{\bf Migration of mesenchymal cells from a cluster.} Initially, $N_m=10$ blue M cells are randomly placed in a circular aggregate with 101 red passive cells with $E_{\theta c}=1$ kPa, $E_{\theta m}=15$ kPa, $j_{cs}=1$, $j_{cc}=1$, $j_{cm}=10$, $j_{ms}=1$, and $j_{mm}=1/2$. Snapshots at MCTS: (a) $t=15$, (b) $t=300$, (c) $t=2700$, and (d) $t=3000$. Attraction point located at the upper-right corner. See Video S3 \cite{suppl}.}
\label{fig8}
\end{figure}

Fig.~\ref{fig8} shows a circular aggregate of 111 cells, with ten randomly placed blue M cells, and 101 red E cells. We observe intrinsic patterns in Figs.~\ref{fig8}(a) and \ref{fig8}(b), such as the elongation of individually migrating cells at the aggregate outer rim, surrounded by round E cells. The M cells escape and migrate effectively toward the attracting point at the upper right corner. Figs.~\ref{fig8}(b)-(d) show other patterns: clusters of two M cells that move forward, break bonds with E cells and deform the outer rim of the aggregate; see Video S3 in \cite{suppl}. Isolated M cells surrounded by E cells are trapped and cannot escape.

Hybrid E/M cells form clusters, migrate collectively. These cells are more adhesive than M cells but retain motile traits, making them effective in metastasis. To model them, we assign them the same energy threshold as M cells, $E_{\theta m}=15$ kPa, whereas E cells have $E_{\theta c}=1$ kPa. In addition, we give hybrid cells specific adhesion parameters: $j_{hs}$, $j_{hc}$, $j_{hm}$, $j_{hh}$ (hybrid-substrate, hybrid-passive, hybrid-motile, and hybrid-hybrid adhesion, respectively). 

\begin{figure}[!h]
\centering
\includegraphics[clip,width=4.3cm]{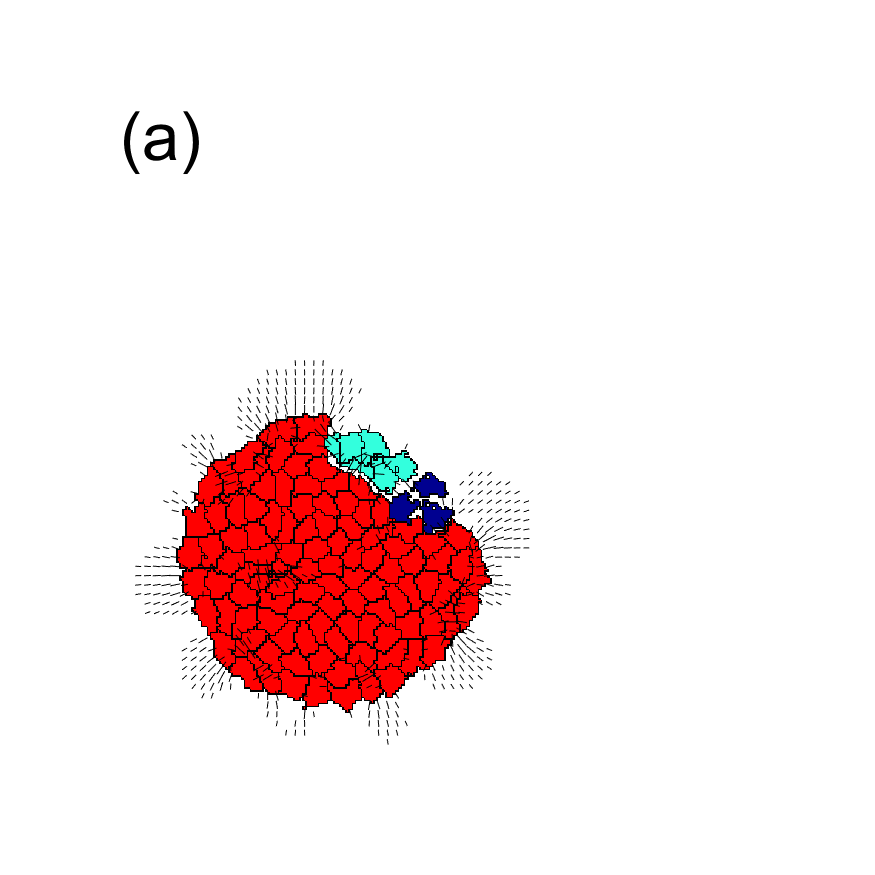}
\includegraphics[clip,width=4.3cm]{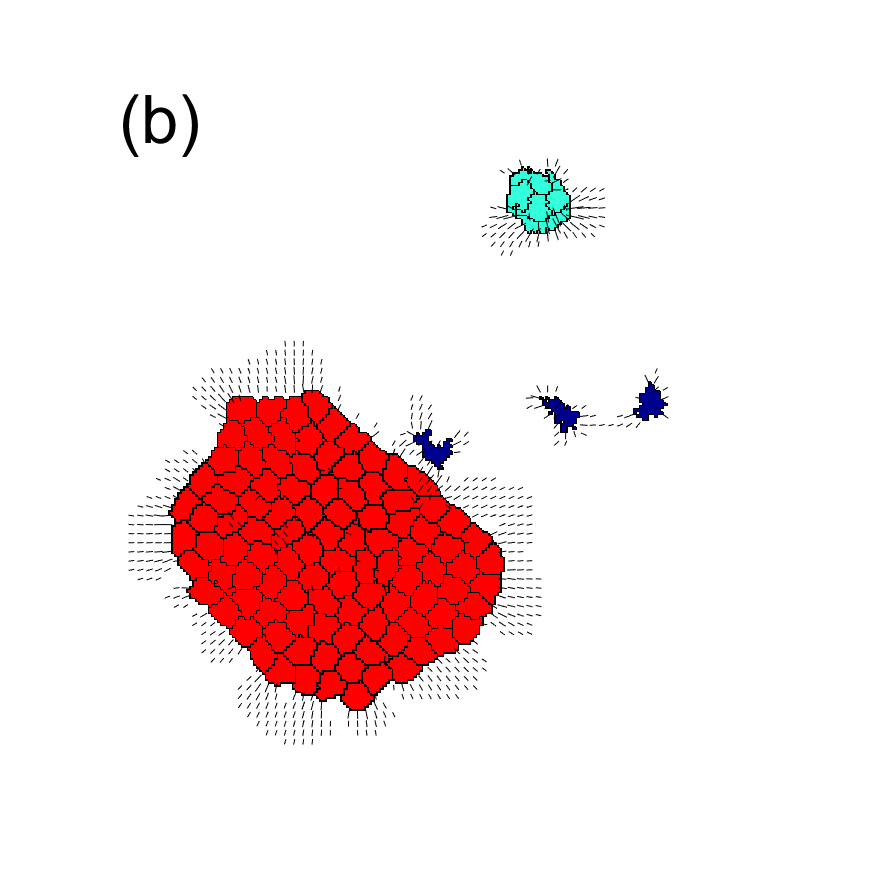}
\includegraphics[clip,width=4.3cm]{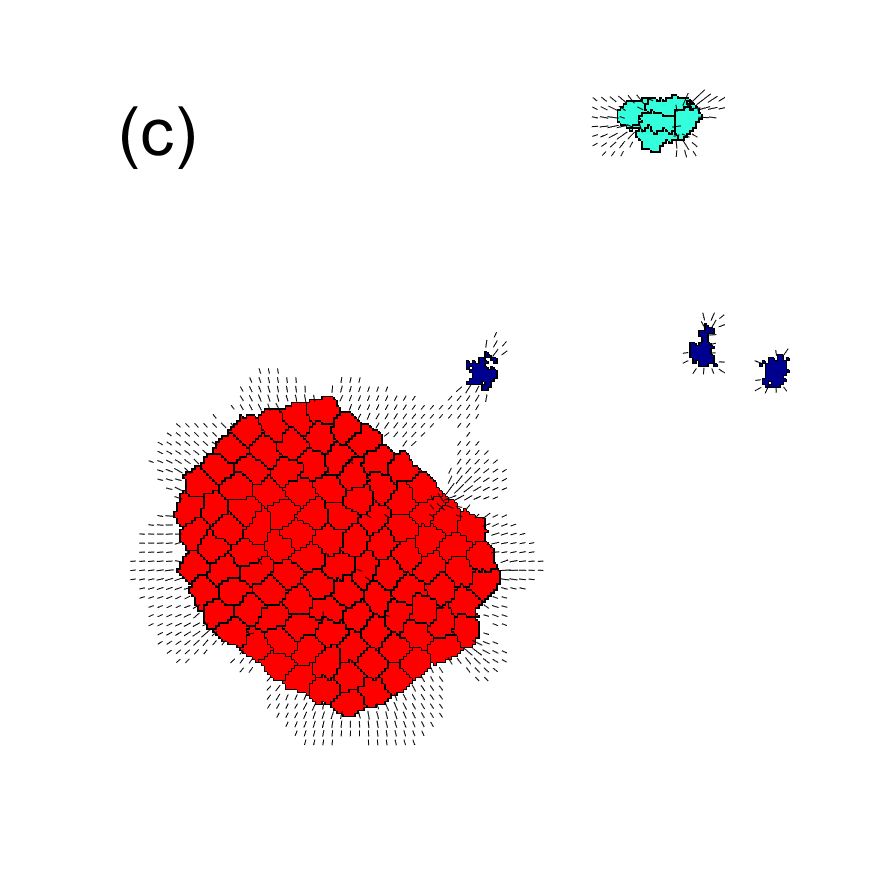}
\caption{{\bf Simulation of adhesive, mesenchymal, and hybrid cells.} $N_{total}=111$ cells comprising $N_m=3$ blue motile M cells, $N_h=4$ light blue hybrid E/M cells, and red passive E cells are randomly placed inside a circular enclosure, with motile and hybrid E/M cells situated closer to the attraction point at the upper-right corner. Parameter configuration: $E_{\theta c}=1$ kPa, $E_{\theta m}=15$ kPa, $j_{cs}=1$, $j_{cc}=1$, $j_{cm}=4$, $j_{ms}=1/4$, $j_{mm}=6$, $j_{hs}=2$, $j_{hc}=10$, $j_{hm}=10$, and $j_{hh}=1/2$. Snapshots at MCTS: (a) $t=20$, (b) $t=1500$, (c) $t=3000$. See  Video S4  \cite{suppl}.}
\label{fig9}
\end{figure}

Fig~\ref{fig9} illustrates three cell types: E (red), M (blue), and E/M hybrid (light blue). M and hybrid E/M cells are placed near the top corner, simulating local influences on phenotype. passive E cells have adhesion parameters that favor attachment to their own type, $j_{cs}=1$, $j_{cc}=1$, M cells have adhesion parameters that encourage them to remain isolated and interact primarily with the substrate, $j_{ms}=1/4$, $j_{cm}=4$, $j_{mm}=6$, $j_{hm}=10$, whereas hybrid E/M cells exhibit a preference for clustering with their own kind but retain motile behavior: $j_{hh}=1/2$, $j_{hs}=2$, $j_{hc}=10$. Fig.~\ref{fig9} shows that $N_h$ hybrid E/M cells form clusters and consistently advance further than $N_m$ M cells within the same time frame, which is consistent with known observations \cite{fri12,vil21}. See Video S4 \cite{suppl} for the complete process.

\begin{figure}[!h]
\centering
\includegraphics[clip,width=4.3cm]{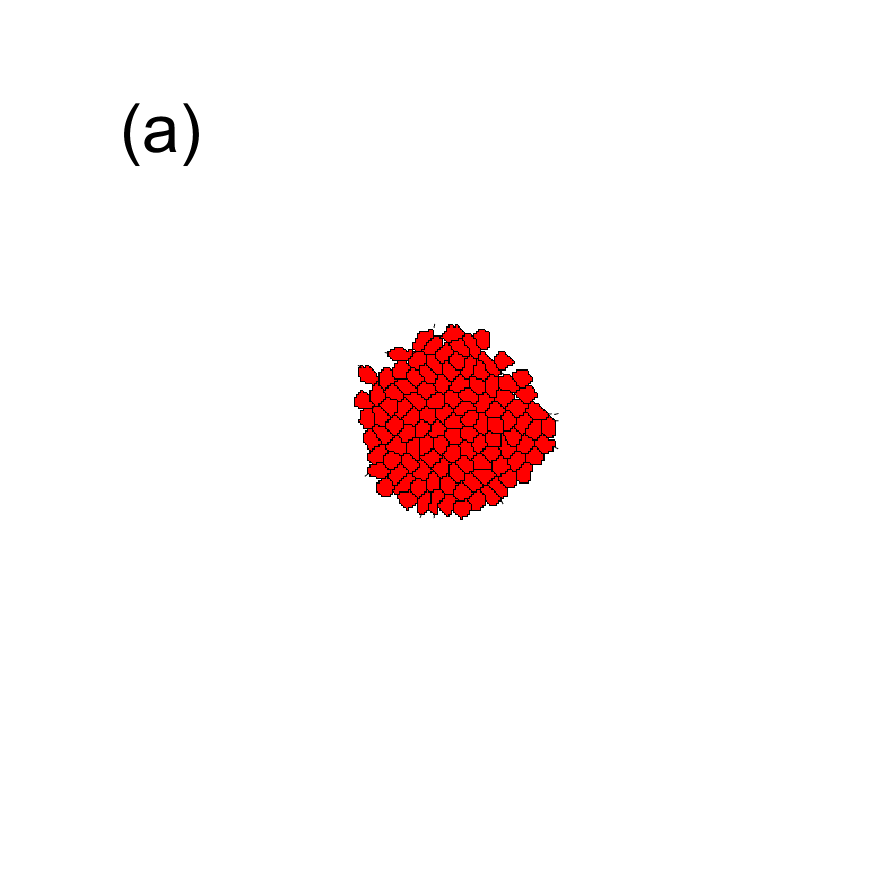}
\includegraphics[clip,width=4.3cm]{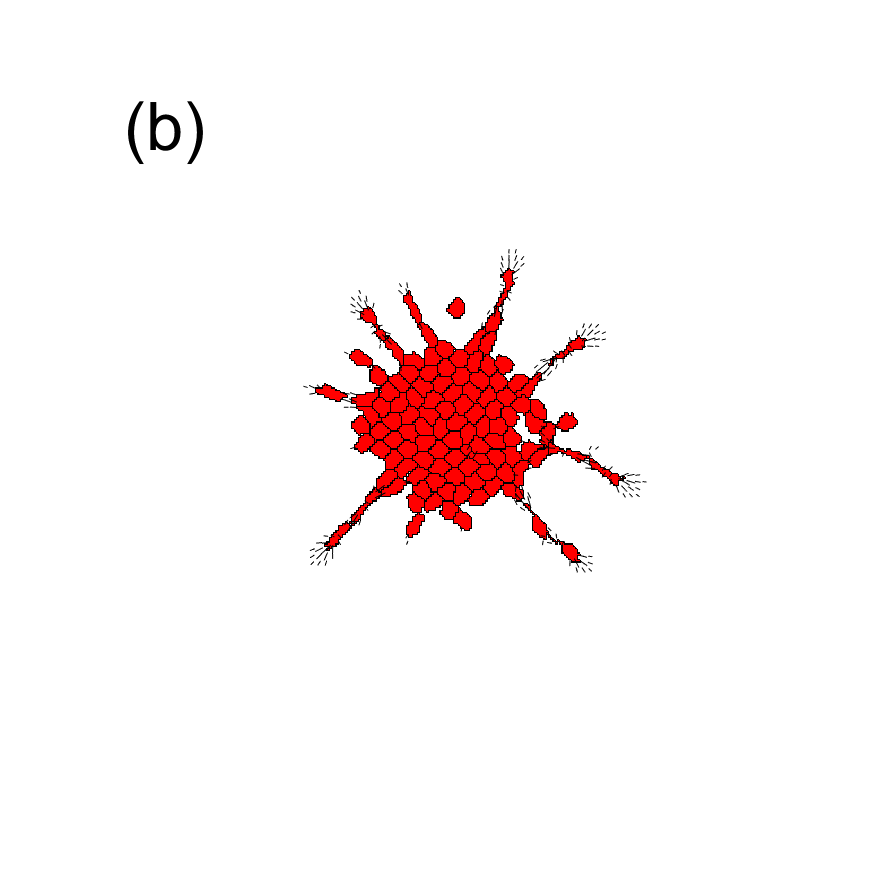}
\includegraphics[clip,width=4.3cm]{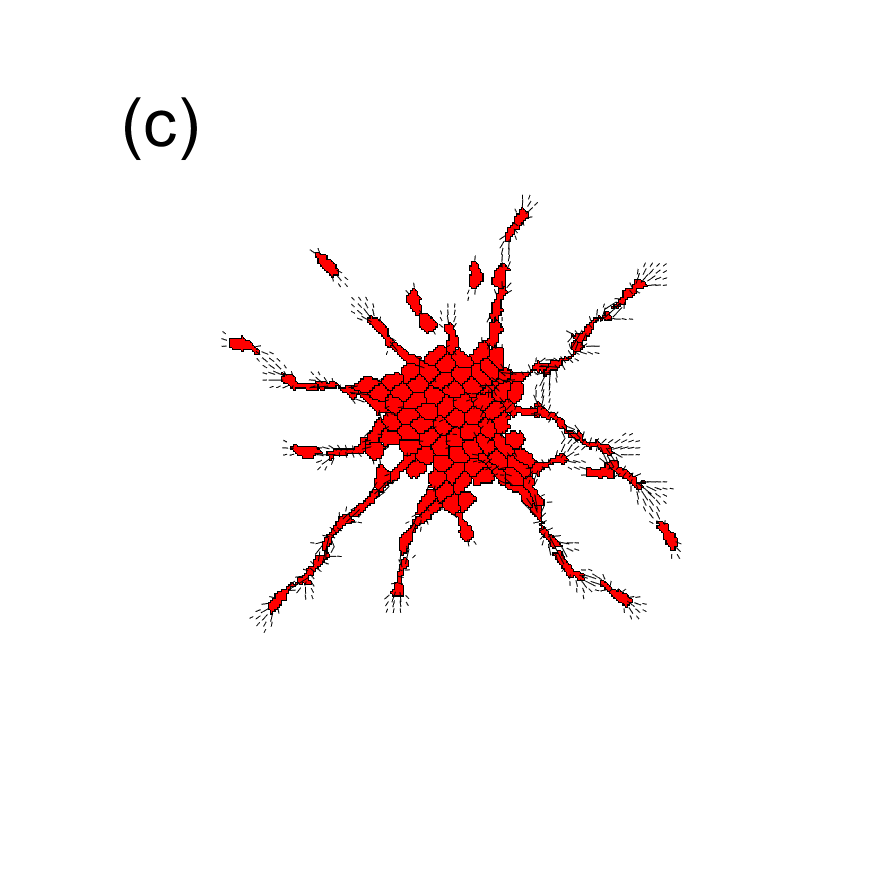}
\caption{{\bf Invasion of tissue from fingering instability of a circular aggregate of epithelial cells.}  Parameter configuration: $N_{total}=111$ cells, $E_{\theta c} =E_{\theta m} = 10$ kPa, $j_{cs}=1$, $j_{cc}=2$. Snapshots at MCTS: (a) $t=30$. (b) $t=500$. (c) $t=1000$. Note that individual cells and small cell clusters detach from the fingers. See Video S5 in \cite{suppl}.}
\label{fig10}
\end{figure}

\begin{figure}[!h]
\centering
\includegraphics[clip,width=4.3cm]{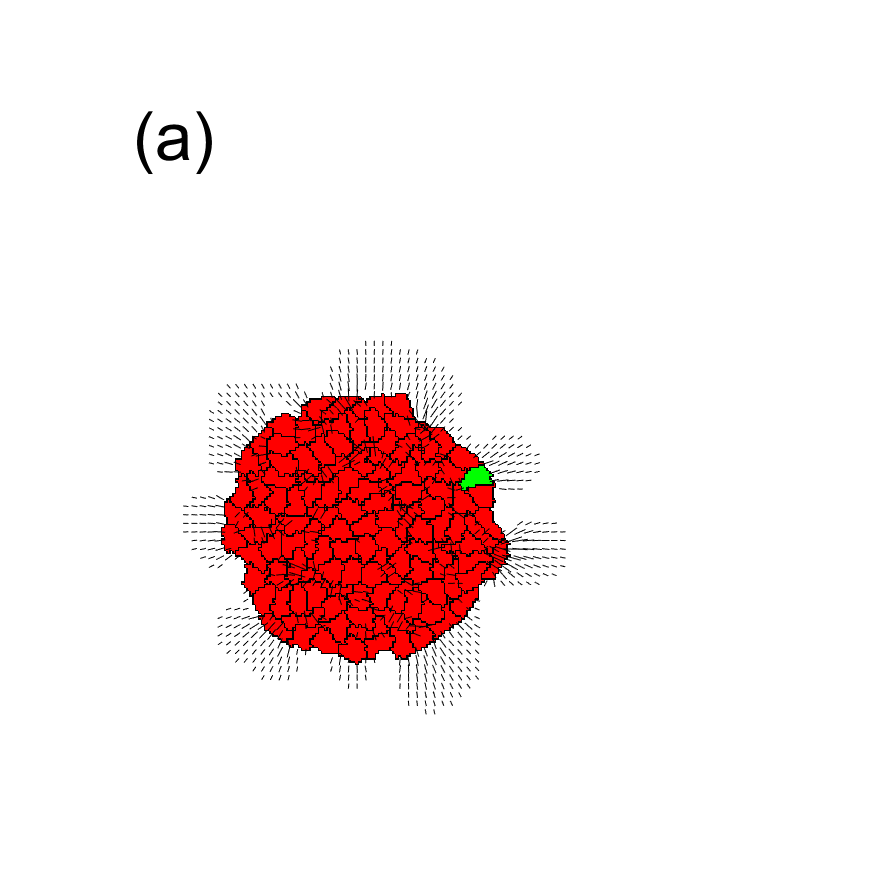}
\includegraphics[clip,width=4.3cm]{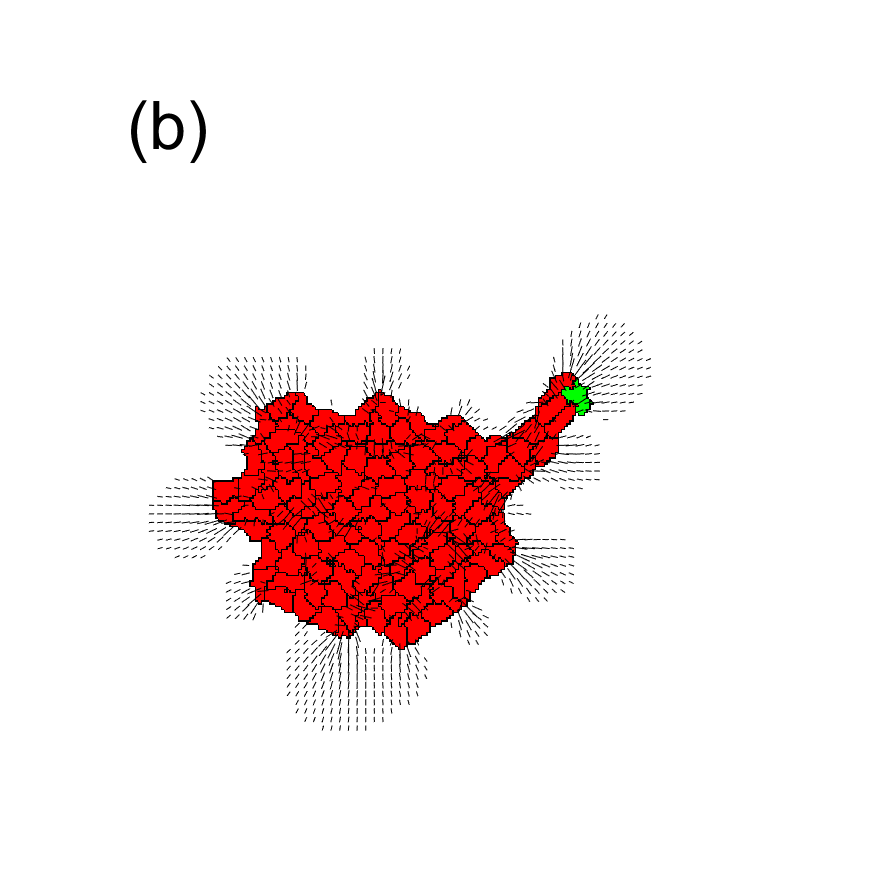}
\includegraphics[clip,width=4.3cm]{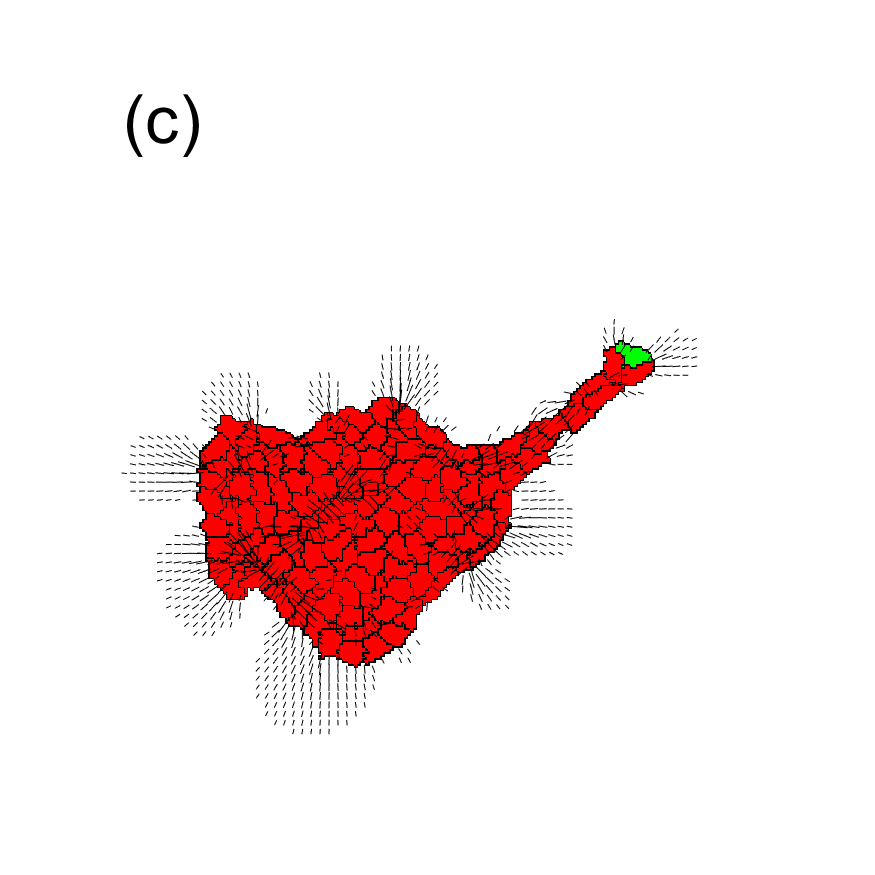}
\includegraphics[clip,width=4.3cm]{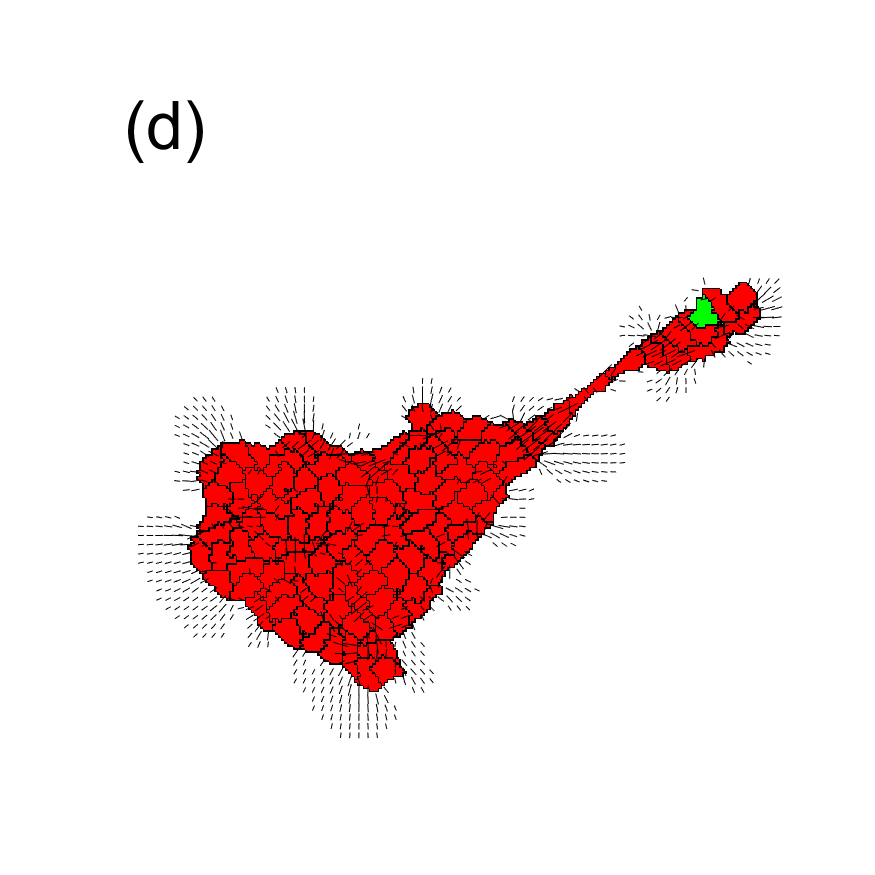}
\includegraphics[clip,width=4.3cm]{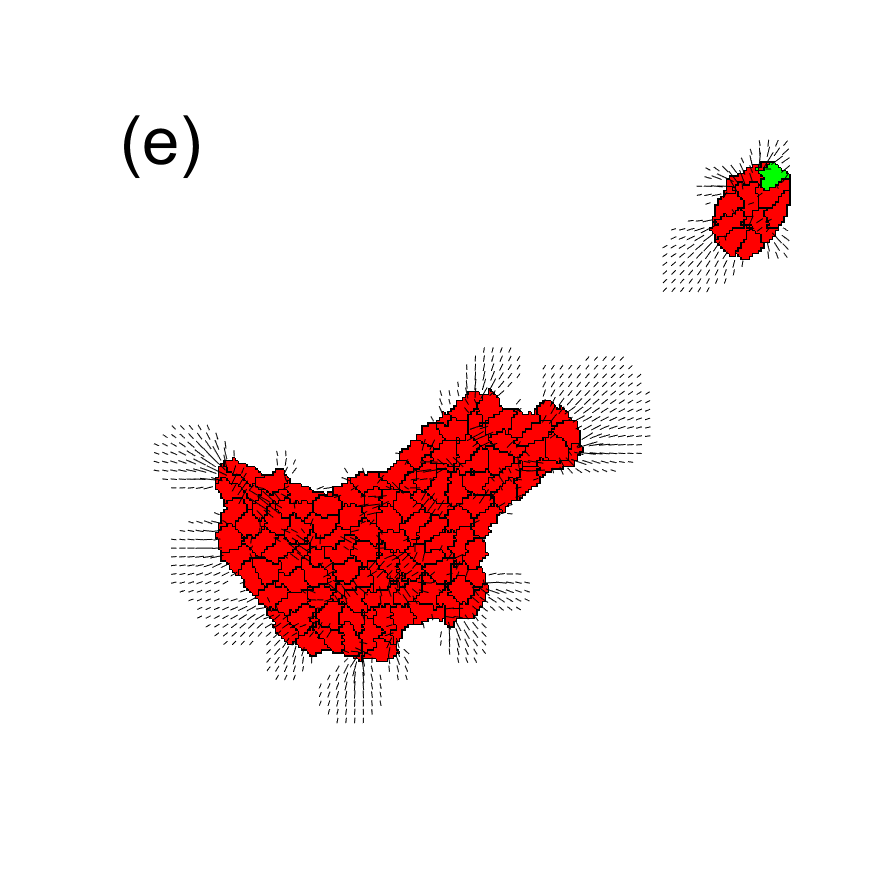}
\caption{{\bf Invasion of tissue from finger formation led by a basal cell.}  Parameter configuration: $N_{total}=111$ cells, $E_{\theta c} = E_{\theta m} = 10$ kPa, $j_{cs}=1$, $j_{cc}=1/2$. Snapshots at MCTS: (a) $t=30$. (b) $t=500$. (c) $t=1000$. (d) $t=2000$. (e) $t=3000$. Note that a small cell cluster led by the green basal cell has detached from the finger. See Video S6 in \cite{suppl}.}
\label{fig11}
\end{figure}

A circular aggregate of epithelial cells can exhibit a fingering instability as in wound healing assays \cite{hak17,pet10}. This has already been captured by the CPM without migrating cells (see Figure 6 of Ref.~\cite{vanOers}); see also Fig.~\ref{fig10}. In the latter figure, the fingers of E cells resulting from the instability of the aggregate collectively invade the surrounding tissue. It is possible to model collective invasion of E cells led by a basal cell which is motile and senses the attraction point, whereas all other E cells are not motile. See Fig.~\ref{fig11}. Another possibility is to modify the adhesion parameter of a single E/M hybrid cell, which is justified as it is known that the EMT transition may exhibit different stages of completion \cite{pas18,bra18,kro19}.  

\section{Discussion}\label{sec:6}

We have proposed a biomechanical model of early invasion of cells issuing from aggregates. Its basis is the cellular Potts model incorporating durotaxis \cite{vanOers}, modified to include active migration forces that affect differently mesenchymal M, epithelial E, and hybrid E/M cells. We are interested in early stages before cellular proliferation. Biochemical drivers such as chemotaxis, cellular Notch signaling, cancer stem cells or the EMT \cite{boc18}, will be left for a subsequent publication. Once the EMT model and the most relevant biochemical descriptions are determined for the particular cancer to study \cite{boc18,vil22,muk22,hir23}, they determine cell phenotypes that have different mechanical characteristics. Our predictions have implications for metastasis but they are qualitative rather than quantitative until supplemented by future work.

As in previous works \cite{vanOers,veg20}, we have assumed that cells act as contractile units resulting in a first moment of area representation for their force distribution \cite{lem10}. The resulting traction forces of the cells are realistic (but they {\em are not} the gradients of the CPM Hamiltonian), the strains in the ECM are calculated via the finite-element method and modify the substrate stiffness while keeping the extension-retraction symmetry \cite{vanOers}. This approach produces cells moving by durotaxis. Adding active forces to traction forces interferes with the extension-retraction symmetry and may impede cell migration in some cases. Here have split each MCTS into two: For half a step, we have adopted the simpler durotaxis description of Ref.~\cite{vanOers} and for the other half step we have used a simple  active forces that direct cells towards an attraction center. Including more realistic active migration forces (e.g., nonlinear push and/or pull forces) is possible. For example, if certain cells move by chemotaxis in the gradient of a continuum field  as in angiogenesis \cite{veg20}, the gradient of the corresponding term in the CPM Hamiltonian (calculated using finite differences, interpolation and smoothing \cite{ren19}) produces active chemotactic forces. These forces could  be incorporated via one half of a fractional time step while the traction forces of the cells comprise the other half step, as in the present paper.  

Our simulations highlight how substrate stiffness, substrate active forces and strain feedback loops drive collective cell behavior. These mechanisms are crucial in tissue morphogenesis, where cells migrate and self-organize into functional structures. The results also emphasize the role of heterogeneity in enabling emergent order in multicellular systems.

We have shown invasion by individual M cells coming from a circular aggregate that may model a cancerous tumor. In the absence of active forces, the M cells surround the tumor, as in Fig~\ref{fig4}(b). This pattern is reminiscent of two-colored cancerous moles in some melanoma types where tissue of light color surrounds the darker nucleus; these tumors may eventually produce metastases \cite{mcc14}. Fig~\ref{fig4}(d) is the visual pattern representing the activation of migration in such a tumor and, maybe, the initiation of metastasis. Our computational model also shows that collective invasion led by clusters of hybrid E/M cells is faster and more efficient than invasion led by M cells; see Fig.~\ref{fig9}. In the future, we plan to incorporate the EMT, Notch signaling and cancer stem cells \cite{boc18} to the present hybrid CPM.
\bigskip












\paragraph*{Acknowledgments}
This work has been supported by the FEDER/Ministerio de Ciencia, Innovaci\'on y Universidades -- Agencia Estatal de Investigaci\'on (MCIN/AEI/10.13039/\-501100011033) grants  PID2020-112796RB-C22 and PID2024-155528OB-C22. 
\nolinenumbers

%
%
%


\appendix
\section{Parameters and dynamic solutions}\label{ap:a}
Here we enumerate 21 configurations and a table with parameter values corresponding to these configurations that produce seven different patterns depicted in Fig.~\ref{fig2}. 
\bigskip


\vspace{-1cm}
\begin{figure}[!h]
\centering
\includegraphics[clip,width=14cm]{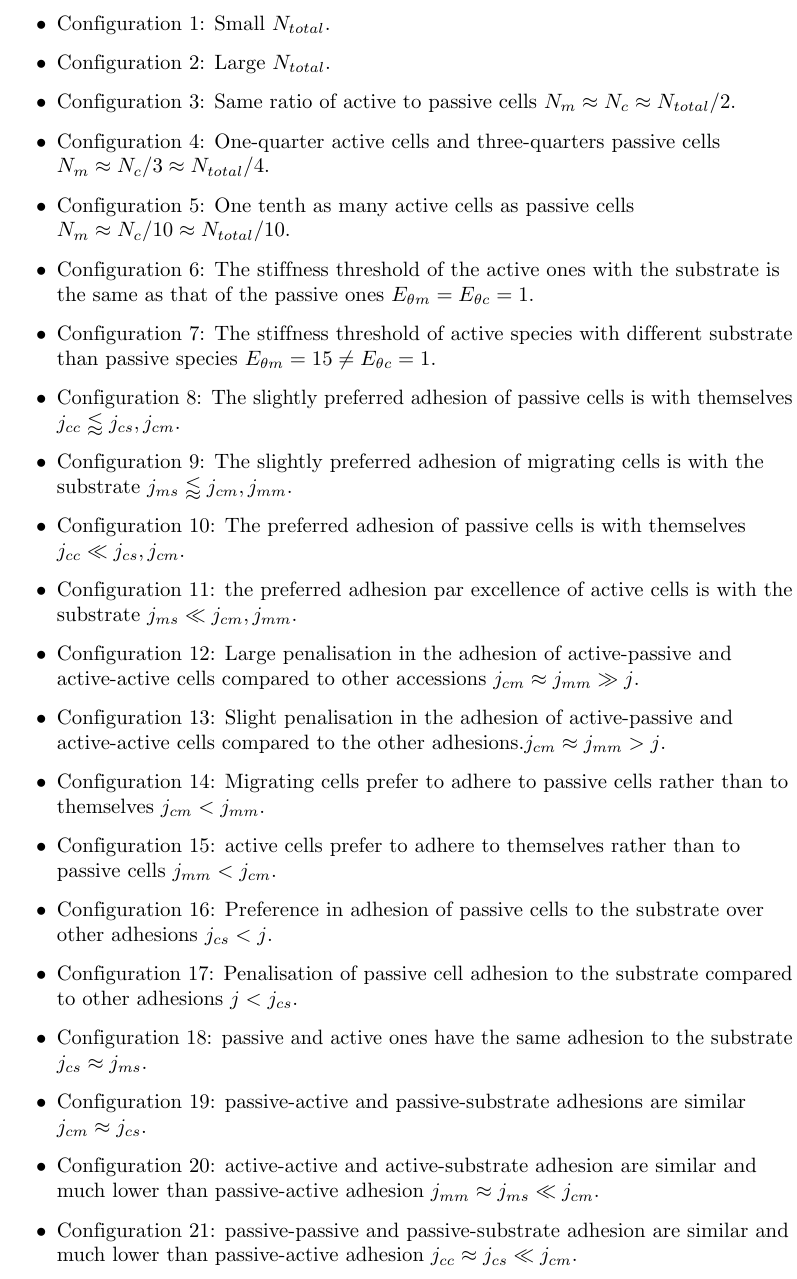}
\end{figure}

\vspace{-0.5cm}
\begin{figure}[!h]
\centering
\includegraphics[clip,width=16cm]{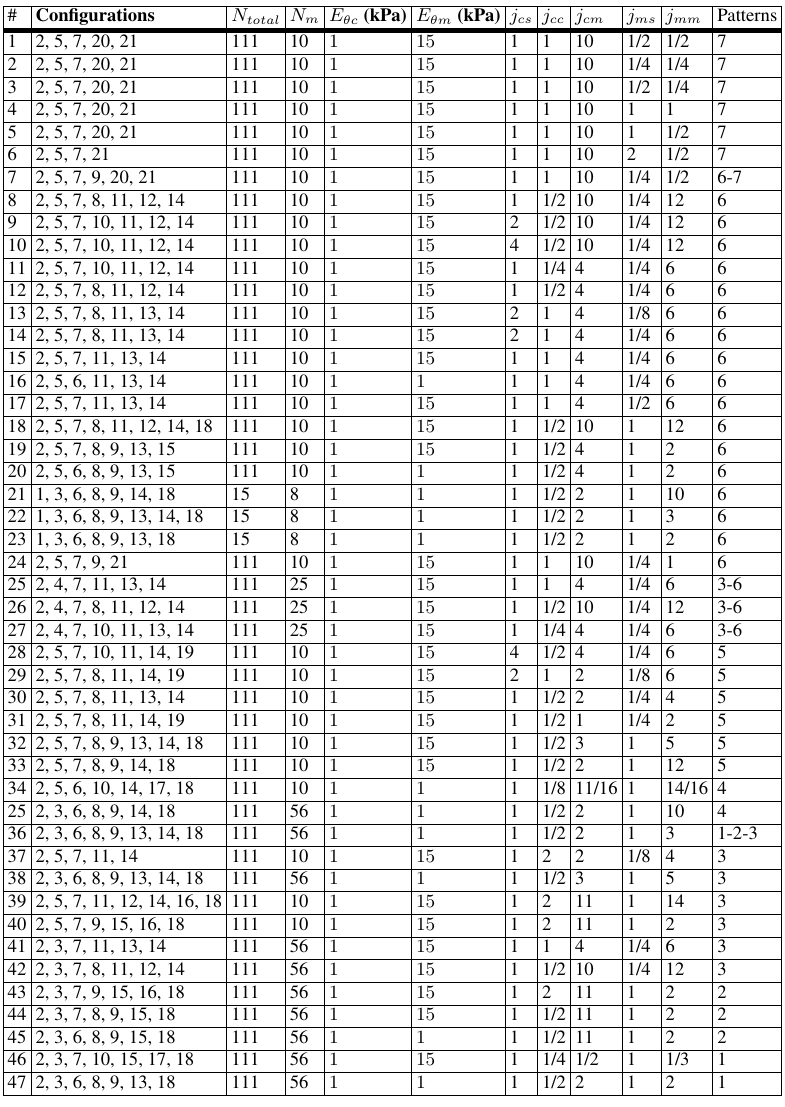}
\caption{}
\label{S1_Table}
\end{figure}

\section{Flowcharts of the fractional step and single step methods}
\label{ap:b}

\begin{center}
\begin{tikzpicture}[node distance=1.8cm]

\node (start) [startstop] {Start: Increment $n$};
\node (FEM1) [processFEM, below of=start] {Compute cell displacement due to ECM stiffness (FEM with force $f$)};
\node (strainH) [process, below of=FEM1] {Compute $\Delta H_\text{duro}$ using displacement and deformation tensor};
\node (CPM1) [process, below of=strainH, text width=10cm, align=center] {CPM moves: $\Delta H = \Delta H_\text{volume} + \Delta H_\text{contact} + \Delta H_\text{duro}$;\\ 
accept/reject via Monte Carlo};
\node (stop1) [startstop, below of=CPM1] {End of first fractional sub-step};
\node (FEM2) [processFEM, below of=stop1] {Compute cell displacement due to ECM migration force $f_m$ (FEM)};
\node (migH) [process, below of=FEM2] {Compute $\Delta H_\text{mig}$ using displacement and deformation tensor};
\node (CPM2) [process, below of=migH, text width=10cm, align=center] {CPM moves: $\Delta H = \Delta H_\text{volume} + \Delta H_\text{contact} + \Delta H_\text{mig}$;\\ 
accept/reject via Monte Carlo};
\node (phenotype) [process, below of=CPM2] {Update cell pixel identities (phenotype assignment)};
\node (stop2) [startstop, below of=phenotype] {End of fractional iteration / ready for next increment};

\draw [arrow] (start) -- (FEM1);
\draw [arrow] (FEM1) -- (strainH);
\draw [arrow] (strainH) -- (CPM1);
\draw [arrow] (CPM1) -- (stop1);
\draw [arrow] (stop1) -- (FEM2);
\draw [arrow] (FEM2) -- (migH);
\draw [arrow] (migH) -- (CPM2);
\draw [arrow] (CPM2) -- (phenotype);
\draw [arrow] (phenotype) -- (stop2);

\draw [arrow] (stop2.east) -- ++(3,0) -- ++(0,16.2) -- (start.east);

\end{tikzpicture}

\end{center}

\vspace{2cm}

\begin{center}
\begin{tikzpicture}[node distance=2cm]

\node (startS) [startstop] {Start: Increment $n$};
\node (FEMS) [processFEM, below of=startS] {Compute cell displacement due to total ECM forces $f + f_m$ (FEM)};
\node (Htot) [process, below of=FEMS] {Compute $\Delta H_\text{duro+mig}$ using displacement and deformation tensor};
\node (CPMS) [process, below of=Htot, text width=10cm, align=center] {CPM moves: $\Delta H = \Delta H_\text{volume} + \Delta H_\text{contact} + \Delta H_\text{duro+mig}$;\\ 
accept/reject via Monte Carlo};
\node (phenotypeS) [process, below of=CPMS] {Update cell pixel identities (phenotype assignment)};
\node (stopS) [startstop, below of=phenotypeS] {End of single step iteration / ready for next increment};

\draw [arrow] (startS) -- (FEMS);
\draw [arrow] (FEMS) -- (Htot);
\draw [arrow] (Htot) -- (CPMS);
\draw [arrow] (CPMS) -- (phenotypeS);
\draw [arrow] (phenotypeS) -- (stopS);

\draw [arrow] (stopS.east) -- ++(3,0) -- ++(0,10) -- (start.east);

\end{tikzpicture}
\end{center}
{\vskip3cm}

\end{document}